\documentclass{emulateapj}
\usepackage{graphicx}
\pdfoutput=1
\usepackage[usenames]{color}

        \renewcommand{\columnsep}{0.5cm}

        \renewcommand{\baselinestretch}{0.97}

\newcommand{\blue}{\color{blue}}

\newcommand{\green}{\color{green}}

\newcommand{\high}{\blue}

\def \etal{{et al.$\,$}}
\def \eg{{\em e.g.}}
\def \gtw{\>\hbox{\lower.25em\hbox{$\buildrel >\over\sim$}}\>}
\def \ltw{\>\hbox{\lower.25em\hbox{$\buildrel <\over\sim$}}\>}

\def \ie{{\em i.e.}}

\def \deg{$^{\circ}$}
\def \arcdeg{^{\circ}}
\def \cm3{cm$^{-3}$}
\def \radm2{ rad/m$^2$}
\def \dyncm2{dyn$~$cm$^{-2}$}
\def \kmps{km$~$s$^{-1}$}
\def \ergps{erg$~$s$^{-1}$}

\def \epsy{\epsilon_{sy}}
\def \hal{H$\alpha$\,}
\def \g{\gamma}
\def \be{\begin{equation}}
\def \ee{\end{equation}}
\shortauthors{Neff, Eilek, \& Owen}
\shorttitle{Cen A Transition Regions: Radio Structure}

\begin{document}

\title{The Complex North Transition Region of Centaurus A: Radio Structure}

\author{Susan G. Neff}
\affil{NASA's Goddard Space Flight Center, 
Laboratory for 
Observational Cosmology,  Mail Code 665,  Greenbelt, Maryland, 20771}
\email{susan.g.neff@nasa.gov}
\author{Jean A. Eilek}
\affil{Physics Department, New Mexico Tech, Socorro NM 87801}
\affil{National Radio Astronomy Observatory\footnote{The 
National Radio Astronomy
Observatory is a facility of the National Science Foundation operated
under cooperative agreement by Associated Universities Inc.}
\footnote{Adjunct Astronomer at the National 
Radio Astronomy Observatory.}
,  Socorro NM 87801}
\author{Frazer N. Owen}
\affil{National Radio Astronomy Observatory$^*$, P. O. Box O,  Socorro NM 87801}
\baselineskip 12pt

\begin{abstract}

    We present deep radio images of the inner $\sim 50$ kpc of 
 Centaurus A, taken with the Karl G. Jansky Very Large Array (VLA) 
 at 90 cm.   We focus on the Transition Regions 
 between the inner galaxy -- including the active nucleus, 
 inner radio lobes and star-forming disk -- and the outer radio 
 lobes.  
We detect previously unknown extended emission around the
Inner Lobes, including radio emission from the star-forming disk.
We find that the radio-loud part of the North Transition Region, known
as the North Middle Lobe,  is significantly
 overpressured relative to the surrounding ISM.  
 We see no evidence for a collimated flow from the Active Galactic Nucleus
 (AGN) through this
 region.  Our images show that the structure identified by Morganti
 \etal (1999) as a possible large-scale jet appears to be part 
 of a narrow
 ridge of emission within the broader, diffuse, radio-loud region.
 This knotty radio ridge is coincident with other striking phenomena:
 compact X-ray knots, ionized gas filaments, and streams of young
 stars. Several short-lived
 phenomena in the North Transition Region, as well as the frequent
 re-energization required by the Outer Lobes, suggest that energy must be
 flowing through both Transition Regions at the present epoch.
 We suggest that the  energy flow is in the form of  a galactic wind.

\end{abstract}

\bigskip

\section{Introduction}

The radio source Centaurus A (``Cen A'') and its parent galaxy,
NGC 5128, form an important active-galaxy system which can reveal
the interplay between the history of the parent galaxy and the
development of the radio source it creates.  Only 3.8 Mpc away,
(Harris \etal 2010;  $1\arcmin \simeq 1.14 $ kpc ),
Cen A/NGC 5128 can be studied with a level of
detail unavailable in other systems.   Interestingly -- or perhaps
because the system is so close -- both the stellar galaxy and the
radio source show details that we do not normally expect.

NGC 5128 is fundamentally a normal elliptical galaxy, 
dominated by an old stellar population 
with kinematic signatures typical of other massive ellipticals. 
The galaxy's optical appearance is, however, somewhat unusual,
being dominated by the iconic central dust band,
which is thought to be the result of a merger with a
smaller gas-rich system. 
 Other remnants of the merger may include cold HI
still orbiting the galaxy and a ribbon of young stars and
emission-line gas clouds  (\eg, Graham 1998, Mould \etal 2000,
Struve \etal 2010).
In a companion paper (Neff, Eilek, \& Owen 2014,
 Paper 2), we show that ribbon extends to $\sim 35$ kpc from the
galaxy, is coincident with a complex, knotty structure seen in radio
and X-rays,   
and is probably energized by a galactic wind moving through the middle
regions of the radio source.

Our focus in this paper is the radio structure of the
inner $\sim 50$ kpc of Cen A.  We  present new 
VLA 90 cm observations,  and discuss their implications 
for the astrophysics of the region and the overall energy
flow in the system.  
In Section \ref{Background}, we review relevant background 
information on the Cen A / NGC 5128 system and present a first look
at our results.
Section \ref{Obs_Imaging} describes our radio observations and 
imaging methods. Section \ref{full_middle_region}
presents our 90 cm image, and discusses its limitations.
  Section \ref{compare_to_M99}
compares our image to the 20 cm results from 
Morganti \etal, 1999 (``M99'').
In Sections \ref{Results_section} and \ref{results_radio_knots}, 
we present and analyze 
various components of the Inner and Transition regions.  
In Section \ref{Energy_Flow}, we discuss the need for energy
flow through the Transition Regions and  argue  that
those regions are dominated by a wind rather than jets.
Section \ref{The_Last_Section} summarizes our results.
\bigskip

\section{Cen A:  Setting the Stage}
\label{Background}

\begin{figure}[htb]
\begin{minipage}[c]{0.475\columnwidth}
\hskip -40pt
\includegraphics[width=2.9in]{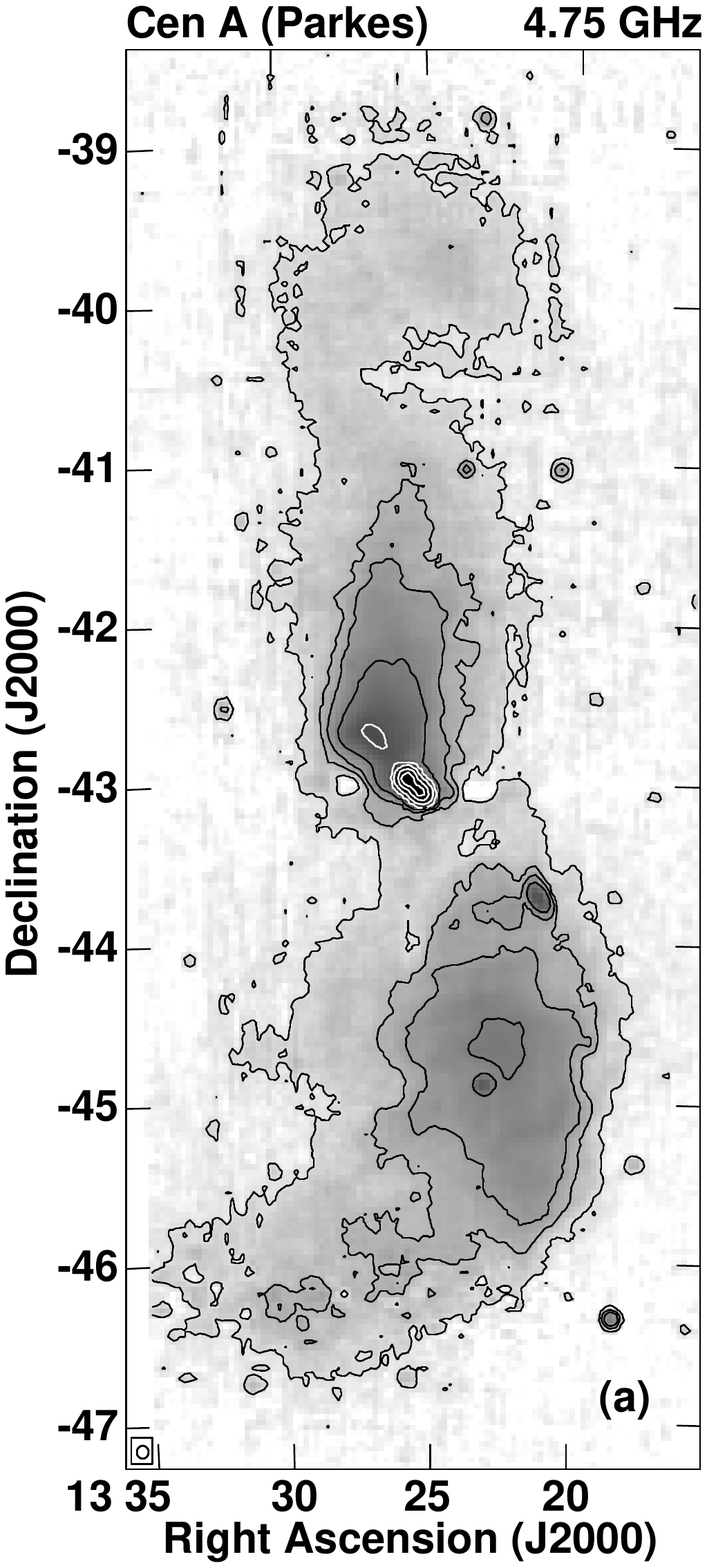}
\end{minipage}
\begin{minipage}[c]{0.485\columnwidth}
\vskip -20pt
\includegraphics[width=1.72in]{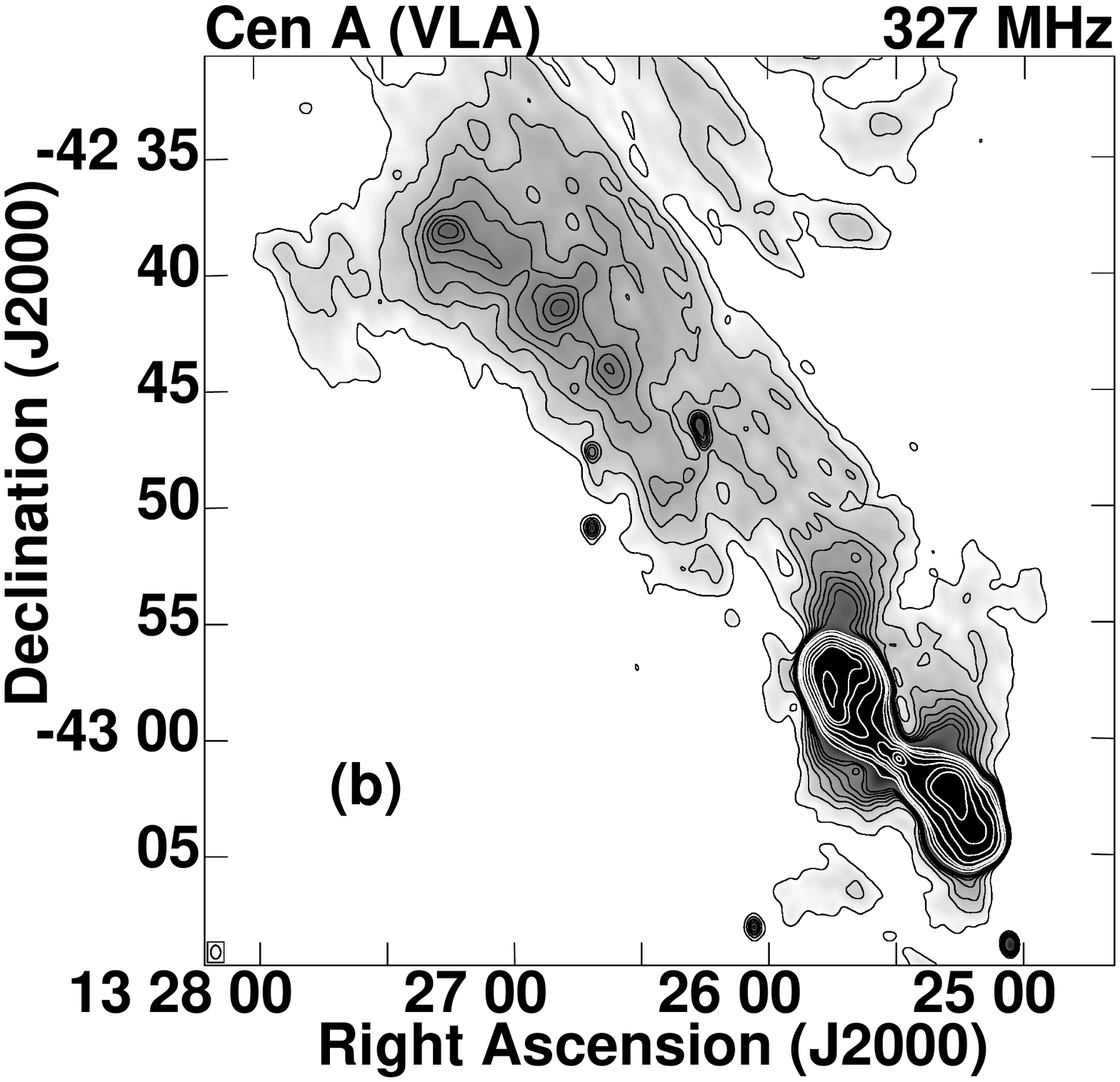}
\vskip -50pt
\includegraphics[width=1.65in]{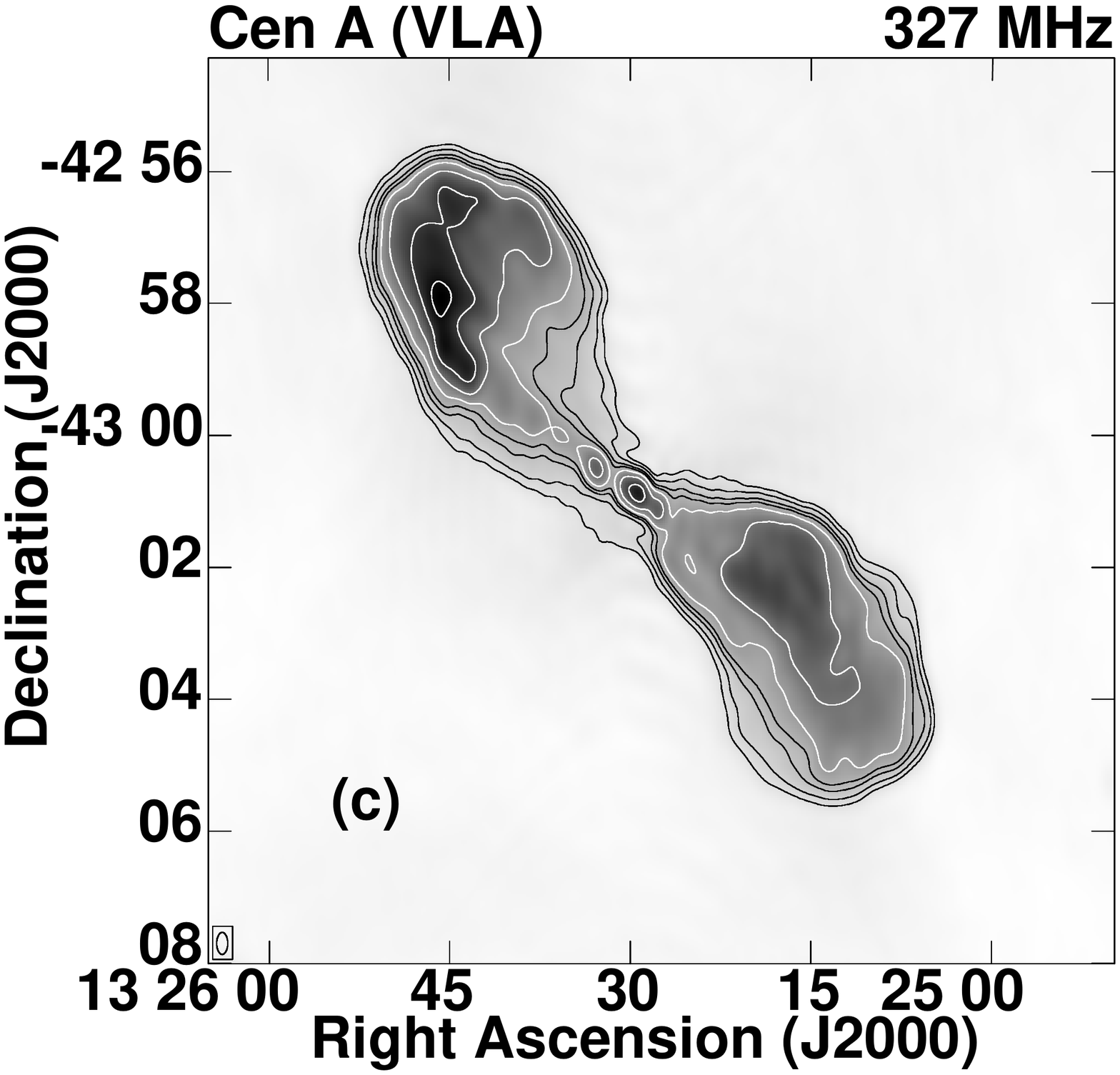}
\end{minipage}
\caption{The three scales of Cen A as seen in radio.  Left (a), the
  large-scale Outer Lobes (also known as ``giant'' lobes); 
5-GHz Parkes image (Junkes \etal~ (1993), kindly provided by N.
Junkes).  Contour  levels 0.1, 0.3, 0.5, 1, 3, 10, 30, 60 
$\times~ 624$ mJy/beam. 
Upper Right (b), our 327-MHz image showing 35 kpc of the
  intermediate-scale North Middle Lobe.  Contour
levels 6, 8, 10, 12, 14, 16, 18, 20, 40, 70, 100, 200, 400, 700, 1000,
2000  $\times~ 10$ mJy/beam. Lower Right (c), our
327-MHz image of the small-scale, 15-kpc Inner Lobes.  Contour
levels 1, 2, 4, 6, 10, 20, 40, 60 ~$\times ~100$ mJy/beam.
 }
\label{Fig:ThreeWay}
\end{figure}

 We begin 
with the three spatial scales of the system -- small, 
large and intermediate, as summarized in Figure \ref{Fig:ThreeWay}
and Table \ref{Acronyms} --
and continue with an introduction to key aspects of the system. 
All distances within Cen A are given in projection on the sky.

 Figure \ref{Fig:ThreeWay}a (from Junkes  \etal 1993),
shows the entire radio source, including the bright Inner Lobes,
the relatively bright North Middle Lobe, 
and the extended Outer Lobes. We note there is no southern counterpart
to the North Middle Lobe.  Figures \ref{Fig:ThreeWay}b and \ref{Fig:ThreeWay}c
 are from this paper. 
Most of our focus in this paper is Figure \ref{Fig:ThreeWay}b:  we
present a new look at the base of the North Middle Lobe.
Our VLA observations show new detail around 
the Inner lobes and the base of the North Middle Lobe,
 leading to a new interpreation of these phenemona. 
We will present newly detected diffuse emission around the Inner Lobes,
including emission from the starburst in the central gas/dust
disk in NGC 5128.  
We see no evidence of the ``large-scale jet'', reported by M99
as extending beyond the North Inner Lobe 
(in Section \ref{compare_to_M99}
 we compare our results to the M99 image).
Because our images beyond the base of the North Middle Lobe
are contaminated by  emission from the outer lobes, our focus 
will be on the region shown in Figure \ref{Fig:ThreeWay}b.

\begin{table}[htb]
\caption{Specialized acronyms related to Cen A}
\label{Acronyms}
\begin{center}
\begin{tabular}{c c c }
\hline
\rule{0pt}{16pt} \hspace{-8pt }
Acronym & Definition &  Scale$^{\rm a}$ \\
  &  & (kpc)
 \\[4pt]
\hline 
\\[-4pt]
 (N,S)ILs & (North, South) Inner radio Lobes  & $\sim7$, $\sim5.5$ 
\\[2pt]
 (N,S)TRs & (North, South) Transition Regions & $10-40$
\\[2pt]
 NML$^{\rm b}$ & North Middle radio Lobe & $10-40$
\\[2pt]
 (N,S)OLs   & (North, South) Outer radio Lobes  & $40-300$ 
\\[2pt]
\hline
\end{tabular}
\end{center}
$^{\rm a}$ Spatial scales give approximate distance from AGN, in projection.
\\
$^{\rm b}$ We use ``North Middle Lobe'' only when referring to radio emission 
from the North Transition Region.\\
\\[10pt]
\end{table}

\subsection{Large scales:  the outer lobes}
\label{Intro:large_scales}

Cen  A is  perhaps best  known  for  its large-scale,  
double-lobed,  FRI  radio source, extending  
$\sim 600$ kpc end-to-end (Figure \ref{Fig:ThreeWay}a, also  Junkes \etal
  1993, Feain \etal 2011).   In addition to being radio-loud,
 the Outer Lobes are 
extended $\gamma$-ray sources  (\eg, Yang \etal 2012).
  The North and South Outer Lobes are generally similar, 
  but differ in detail.  The North Outer Lobe is straighter and 
  extends $\sim 4.5^{\circ}$ ($\sim 270\arcmin$) on the sky, 
  while the South Outer Lobe bends from south to east and 
  extends $\sim 4^{\circ}$ ($\sim 240\arcmin$).

  The Outer Lobes are relatively old.  
  Eilek (2014; ``E14'') argues, based on dynamical models, 
  that the age of the large-scale radio structure is
  on the order of 1 Gyr and is no younger than $\sim 400$ Myr old. 
  Because this age is much
  greater than the radiative lifetimes of electrons in the Outer Lobes (at
  most $\sim 80$ Myr for radio-loud electrons, and only $\sim 1$ Myr 
  for the
  electrons creating the $\g$-rays by inverse Compton scattering of
  the Cosmic Microwave Background Radiation; \eg, E14), 
  it is clear that the outer lobes are not
  ``dead''.  Rather, they must have been energized quite recently in
  order to keep shining, especially in $\g$-rays.  E14 argues that MHD
  turbulence within the lobes can provide the necessary {\it in situ}
  energization, but that the turbulence itself will decay in no more
  than $\sim 30$ Myr unless it is resupplied from the Active Galactic
  Nucleus (AGN).  Thus, the
  outer lobes cannot have been disconnected from any driver for more
  than a few tens of Myr.

\subsection{Small scales:  the AGN and the central galaxy}
\label{Intro:small_scales}

The inner few kpc of the galaxy contains two sources of power,
both of which are 
quite young compared to the $\sim$Gyr age of the outer radio lobes.

The massive black hole in the nucleus of NGC 5128 is active right
now, driving out pc-scale jets (\eg,  Muller \etal 2011). 
Plasma from the nuclear jets accumulates in two inner radio lobes, 
which extend $\sim7$ kpc NE and $\sim5.5$ kpc SW of the
radio core (Figure \ref{Fig:ThreeWay}c).  The northern 
jet can be traced continuously in radio (Burns \etal 1983) 
and X-rays (Schreier \etal 1981) from the galactic core 
out to $\sim5$ kpc, where it bends, loses collimation, and 
expands to create the North Inner Lobe (NIL).  
No kpc-scale jet has been found in the South Inner Lobe 
(SIL), although a few compact knots SW of the core  may 
indicate an unseen jet 
(Goodger \etal 2009), and filamentary structure within 
the SIL suggests an ordered flow. 

X-ray observations show that NGC 5128's interstellar medium (ISM) is
disturbed by, and interacting with, the radio jet and Inner Lobes. Two
X-ray bright patches (Kraft \etal 2007) suggest an interaction of the
North Inner Lobe with the local ISM, and a discontinuity in X-ray
emission appears to coincide with the jet decollimation (Kraft \etal
2008). To the south, X-ray images show the expanding South Inner Lobe
is driving a shock into the ISM.  Because the shock strength can be
related to the expansion speed of the South Inner Lobe, 
these data show that the
Inner Lobes are only a few Myr old (Section
\ref{Energy_throughput}, also Croston \etal 2009).

A starburst is currently
underway in the core of NGC 5128. The  central, warped gas and dust lane
contains young stars (\eg,  Dufour \etal 1979), 
young clusters (Minitti \etal 2004), and HII regions 
(Bland \etal 1987), all of which
 indicate ongoing active star formation in the disk.
This starburst has continued for at least 50 Myr,
and probably longer.  In Paper 2 we argue the
starburst and AGN together are driving a wind which carries energy
into and through the middle regions of the radio source.

\subsection{Intermediate scales:  the transition regions}
\label{TransRegion_overview}

Our focus in this paper is the regions between the inner and outer radio 
lobes.    We refer to these regions, $\sim$10$-$40 kpc 
north and south of the galaxy, as 
 the North and South Transition Regions (NTR, STR). 


\paragraph{Radio emission in the transition regions}
To the north, low-resolution radio images (\eg,
Figure \ref{Fig:ThreeWay}a, from Junkes \etal
1993; also Hardcastle \etal 2009)
 show a conspicuous bright structure that appears to connect
the north end of the North Inner Lobe to the inner part of the North
Outer Lobe.\footnote{The North Middle Lobe can also be detected in the
un-enhanced image from Feain \etal 2011 (their Figure 3, see also
Figure 1 of Eilek 2014 for a different stretch).}
This structure extends to at least  $\sim 40$ kpc from the
active core of NGC 5128 and is the region of highest radio surface
brightness after the inner lobes.  It is situated where the
larger-scale radio source gradually bends from the NE orientation of
the northern jet and North Inner Lobe, towards the NW and 
thence into the North Outer Lobe.  This structure is known as the
North Middle Lobe in
the literature, and we follow that tradition in this paper when we
refer to the radio-loud part of the North Transition Region. 
Figure \ref{Fig:ThreeWay}b  shows
our new 90-cm image, which we discuss in detail in 
Section \ref{Results_section}.

No comparable feature exists to the south;  there is no ``South Middle
Lobe''.  Junkes \etal (1993) detect
a diffuse bridge of faint radio emission connecting the South Outer
Lobe to the core and South Inner Lobe, $\sim 10$ times fainter surface
brightness than the corresponding region to the north.  This diffuse
emission is resolved out in higher-resolution radio images (\eg, Feain
\etal 2011, also our new images in Section \ref{Results_section}).

M99 explored the North Middle Lobe 
 at higher resolution {\bf($56\arcsec \times 30\arcsec$)}.  
Their 1.4-GHz image shows a quasi-linear structure, which starts at 
the outer edge of the North Inner lobe,  
 extends  $\sim 10$ kpc to the NE,
and then broadens out to make up part of
the diffuse structure known as the 
North Middle Lobe.   M99 referred to the
linear structure as a ``large-scale jet'', to distinguish it from the
few-kpc inner jet, but also emphasized that its true nature 
is unclear. 
We discuss this linear feature, which we do not see in our images, 
at length from an imaging point of view in Section \ref{compare_to_M99} and 
from the point of view of the energy flow in Section \ref{Energy_flow_in_TRs}.

    
\paragraph{X-ray-loud ISM}  

The quiescent ISM of NGC 5128 can be traced to $\sim 30$ kpc from the
galaxy, both to the NW and the SE (\eg, Feigelson \etal 1981, Kraft
\etal 2009; ``K09'').  An X-ray ``hole'' north of 
the galaxy (Kraft, 2009) appears to be coincident with the NML, 
suggesting that radio-loud plasma has created a cavity in 
the ISM of NGC 5128 similar to those seen in other radio galaxies.  
Feigelson \etal
(1981) found an X-ray ridge near the SE edge of the North Middle
Lobe, close to the apparent boundary between the hot, thermal ISM of
the galaxy (SE of the ridge) and the nonthermal plasma of the NML (NW
of the ridge).  XMM-Newton observations (K09) show that the X-ray
ridge contains several compact, X-ray-loud knots.  In Section 
\ref{NML_Xrayknots} we show this knotty X-ray
ridge is closely related to the radio-bright ridge 
 we describe in Section \ref{Results_section}.


\paragraph{``Weather'' in the north transition region}
In addition to the unusual radio  and X-ray emission,
the North Transition Region contains a string of 
optical emission-line filaments and recently formed stars 
(Graham \& Price 1981, Graham 1998, Mould \etal 2000, 
Crockett \etal 2012).  In Paper 2 we show that
the emission line region extends to $\sim 35$ kpc from the
center of the galaxy, and this 
structure is also detected in \hal and Far-UV emission.
By analogy to terrestrial weather, where
memorable events (\eg,  tornadoes and cyclones) can be dramatic local
consequences of atmospheric flows, we refer to this complex mixture
of radio and X-ray emission, disturbed multiphase ISM, and star formation as
the {\it ``weather system.''}  We explore this
system  in greater detail in Paper 2.


\subsection{Models for the transition regions}
\label{NML_models}

The astrophysics of the transition regions is not well  understood.
 The radio loudness of the North Middle Lobe and the emission line
filaments in the North Transition Region have received the most 
attention to date.

\paragraph{Radio loudness of the North Middle Lobe}
Although a number of models have been advanced to explain the 
radio-loudness and structure  of the NML, 
none has emerged as definitive, 
and none explains all aspects of the region.  

Several authors have argued that the linear feature seen by M99
is, indeed, a steady jet 
which extends past the northern edge of the 
North Inner Lobe
and creates the North Middle Lobe.
Junkes \etal (1993) suggested the NML is the region where a large-scale
jet first loses energy and creates an expanding plasma cloud.  
Romero \etal (1996) suggested the NML is a ``working surface'' where 
a supersonic jet bends as it runs into the local ISM. 
Gopal-Krishna and Wiita (2010) suggested that the NML results from an
interaction between a jet  and gas clouds associated with a stellar shell.
All of these suggestions are challenged by our lack of
detection (at 90 cm) of a narrow collimated jet 
within the NML region.

Another common theme holds that the North Middle Lobe was 
created by a time-variable 
active nucleus.  Saxton \etal (2001) model the NML 
as a buoyant plasma bubble, formed by a previous 
outburst of the central radio jet, which has pulled
a stream of plasma from the north inner lobe as it rises.
Similarly, K09 argue that a bubble was
created in an earlier activity cycle, and the
bubble is now being energized by plasma
outflow from the current AGN outburst and thus seen as the radio-loud
NML. These models are challenged by the disparity between the long time
necessary for the slowly-rising bouyant bubble to reach
the North Transition Region ($\sim 140$  Myr, Saxton \etal 2001)
and the many short-lived
features seen in the region 
(as we describe in Section \ref{Short_lived_NTR}).

\paragraph{The ``missing'' South Middle Lobe}
We have found surprisingly little discussion of the north-south asymmetry
of  the middle-scale regions. Haynes \etal (1983) pointed out
that relativistic beaming could cause a radio asymmetry, but did not comment
on why such beaming would not also affect the inner lobes.  
Hardcastle \etal (2009) argued that the absence of a bright South Middle Lobe 
means there is currently no energy flow into the South Outer Lobe.  This
argument is challenged, however, by the  ongoing re-energization
required to keep the South Outer Lobe shining in $\g$-rays and 
high-frequency radio emission
(\eg, E14).  We note that
a useful model of the source must explain the brightness
difference between the North and South Transition Regions, 
as well as clarify other features in the northern weather system.

\paragraph{Star formation and emission line filaments}
There is a general consensus in the literature that an outflow,
usually presumed to be from the AGN,  
is causally related to the emission-line clouds and young stars 
in the weather system.   Several authors have
argued that the narrow feature seen by M99 is a collimated jet which
has energized the emission-line clouds and/or induced star formation
within the NE weather system.   Models include weak shocks in dense clouds 
photoionizing the ambient gas (Sutherland \etal 1993), or shocks
causing cloud fragmentation and collapse with subsequent star formation 
(\eg, Fragile \etal 2004, Croft \etal 2006). 
These models for the weather system 
are challenged by the apparent lack of a jet in the region, as 
well as by substantial distance offsets of the emission line features
from the putative outer jet. 
However, we will argue  that many 
effects attributed to a jet-cloud interaction can equally well 
be caused by a wind-cloud interaction. 
In Paper 2 we return to the galactic wind and explore its likely 
connection to weather in the North Transition Region, and 
to differences between the North and South
Transition Regions.

\bigskip


\section{Observations, Reduction and Imaging}
\label{Obs_Imaging} 

Motivated by the M99  results, we wanted to obtain
a better image of the linear feature in their image,
which was  suggested as a possible large-scale jet. 
The relatively faint radio emission outside the inner double is 
difficult to image at 20 cm or shorter wavelengths, because 
of the extremely bright inner lobes and core.  We therefore 
observed at 90 cm, where
we expected the diffuse emission and possible outer jet to be brighter
and the central double to be less dominant.  

\begin{table}[htb]
\caption{Observing Runs Summary}
\label{Table:Obs}
\begin{center}
\begin{tabular}{c c c}
\hline
\rule{0pt}{16pt} \hspace{-12pt} 
Configuration & Date & Time on-source \\
& & (hours) \\
\hline
BnA   & 2005 January 23  & 3.28\\
BnA   & 2005 January 25  & 3.28 \\
BnA   & 2005 February 03 & 3.31\\
BnA   & 2005 February 12 & 3.27\\
CnB   & 2005 June 21     & 3.28\\
\hline
\end{tabular}
\end{center}
\end{table}

\begin{table*}[htb]
\caption{Images of Centaurus A at 90 cm }
\label{Obs}
\begin{center}
\begin{tabular}{lcccc l}
\hline
\label{Images}
Image & Restoring beam & total flux & rms & Figures & Note \\[2pt]
& B$_{maj} \times$ B$_{min} \times$ P.A. & (Jy)&  ( mJy/beam) & \\[2pt]
\hline
Middle Lobes:   & & & && \\
$\quad$ combined model   &   $28\arcsec  \times  17\arcsec  \times 11\arcdeg$  
      & 1389 & 3.0   &  ---
& (a)  \\[2pt]
$\quad$ convolve35   &   $35\arcsec  \times  25\arcsec  \times 0\arcdeg$  
      & 1389 & 5.3    
  & \ref{Fig:best_overall},\ref{Fig:schematic_overall}
\\[2pt]
$\quad$ convolve40  &   $40\arcsec  \times  30\arcsec  \times 0\arcdeg$  
      & 1389 & 7.2    &  ---     & (c)  \\[2pt]
High Pass Filtered & $40.0\arcsec  \times 30.0\arcsec  \times 0.0\arcdeg$
  & 723 & --       &  \ref{Fig:radio_knots}, \ref{Fig:Radio_Xray}
  &  (d) \\[2pt]

Inner Double     &  &&&& \\
$\quad$ combined model  &  $21.5\arcsec  \times  11.4\arcsec \times 3.5\arcdeg$ 
         & 991 &  3.7  &  ---  & (a) \\[2pt]
$\quad$ convolve22  &  $22\arcsec  \times  12\arcsec \times 0\arcdeg$ 
         & 991 &  4.4  &  \ref{Fig:InnerLobes}(a)  & (b) \\[2pt]
$\quad$ convolve25  &  $25\arcsec  \times  20\arcsec \times 0\arcdeg$ 
         & 991 &  8.0 &  \ref{Fig:InnerLobes}(b)  & (b) \\[2pt]
ATCA 21 cm  & $60.5\arcsec  \times 36.25\arcsec  \times 2.88\arcdeg$
  & 46  & 5.4       &  \ref{Fig:NML_compare}, \ref{Fig:NML_compare2}
        &  (e) \\[2pt]
Matched VLA 90 cm   & $60.5\arcsec  \times 36.25\arcsec  \times 2.88\arcdeg$
  & 624  & 28      &  \ref{Fig:NML_compare},  \ref{Fig:NML_compare2}
        &  (e) \\[2pt]
\hline
\end{tabular}
\label{ImageTable}
\end{center}
(a)  Image is weighted average of images made for each of the 4 
band/polarization combined-day datasets.
\\
(b) Restoring beams used for
images and measurements are indicated in captions.
\\
(c) Image used as starting point for High-pass filtered image.
\\
(d)  Open-filtered image, made from image with restoring beam 
$40\arcsec \times 30\arcsec \times 0\arcdeg$,
to highlight structure on scales less than $\sim 2.7\arcmin$.
\\
(e)  ATCA
image kindly provided by R. Morganti. Matched VLA 90 cm image convolved
to identical resolution.   Neither of these images contain all
of the inner double source;  total fluxes indicate what is in the images and 
do not measure the source's total flux density.
\\[10pt]
\end{table*}


We observed with the VLA in BnA, and CnB configurations for
a total of 16.4 hours on-source between January and June 2005, as
shown in Table  \ref{Table:Obs}.  We observed mainly at night, to avoid
ionospheric phase destabilization and to minimize radio interference from
terrestrial orgins. 
Observations were centered on the active nucleus, at J2000.0 coordinates
13$^{\rm h}$25$^{\rm m}$27.6$^{\rm s}$, -43\deg01\arcmin09\arcsec.  
The VLA was tuned to two bands 
centered at 321.5 and 327.5 MHz;
each band received both left-circular and right circular polarization, giving
us a total of four band/polarization (``band/pol'') pairs. 
Data from each band/pol pair
were collected in spectral line mode, with 15 channels
covering a bandwidth of 5.86 MHz, for a total bandwidth of 
11.7 MHz.
The narrow individual channel observations helped avoid bandwidth
smearing of regions away from the field center 
(beam smearing increases linearly as a function of distance from the 
field center and at $30\arcmin$ from the center 
is $\sim2\arcsec$, very much less than our synthesized beam), 
and also proved helpful in excising Radio Frequency Interference (RFI).

\subsection{Calibration \& Editing}
\label{calibration}

All reductions were done using the Astronomical Imaging Processing 
System (AIPS) package (Greisen 2003). 
For each day's observation, we first split a bright point source
calibrator from the raw database and applied a phase self-calibration.
Then we used this otherwise uncalibrated source to calculate a
bandpass correction which was used to flatten the spectral response.
After initial examination of the data, one channel in the lower band
was deleted 
from all data because of persistent strong RFI.   The data were 
calibrated in the standard way, using the AIPS task SETJY
with 3C286, and assuming the VLA-determined 1999.2 flux 
scale.\footnote{The 1999.2 flux scale gives flux densities of 
26.039 Jy and 25.901 Jy for the flux calibrator 3C286.
The current best-value fluxes for 3C286 at these wavelengths
are 26.20Jy and 26.02 Jy (R. Perley, private communication,
 May 2014).  If the newest values are correct, 
fluxes presented in this paper may be $\sim 0.5\%$ high.}  
An initial data flagging pass was run to 
remove obvious RFI. 

\subsection{Imaging \& Self-calibration} 
\label{Imaging} 

Each of the four combinations of band and polarization, on each of
the five observation days,
was treated separately.  We made three or four passes through each data
set, with a sequence of imaging (CLEAN algorithm), flagging for RFI, 
and self-calibration.  For these iterations, we imaged a single
field $68\arcmin$ across.

Because radio emission from the North Middle Lobe is 
very low surface brightness, we
were concerned that our imaging might be corrupted by nearby sources
in the sidelobes of the primary beam.  
To counter this possible problem, we subtracted 
our source model for Cen A from the visibility data, and  
made a large mosaiced image
extending to $4^{\circ}$ from the pointing center.  
(For comparison, the FWHM of the primary beam of a VLA antenna 
is $\sim 2.25\arcdeg$ at 90 cm, or a radius of $\sim 1.1 \arcdeg$).   
We CLEANed each facet of  this mosaic to remove all 
small-scale sources outside 
the central $\sim 17 \arcmin$.  
A bright radio galaxy, MRC1318-434B (at $13^{\rm h}21^{\rm m}15^{\rm s}$,
 -43\deg41\arcmin21\arcsec) was imaged and removed from the 
visibility data; other sources were faint and most were unresolved.
After the confusing sources were subtracted, the Cen A source model 
was added back into the visibility data.
A few compact sources remained in the central field, projected
on or embedded in the Cen A radio emission, and are visible in our 
images. 

The complicated structure with the bright inner lobes 
proved too complex for a good restoration with 
the clean algorithm, as we have found for other complex sources such as M87.
We therefore moved to the Maximum Entropy (MEM) 
algorithm implemented in the AIPS program VTESS (Cornwell \& Evans, 1985),
in order to produce a higher-fidelity image of the diffuse emission
from the NML.


\begin{figure*}[ht]
{\center
\includegraphics[width=5.in, trim=.3cm .3cm .3cm .3cm, clip=True]{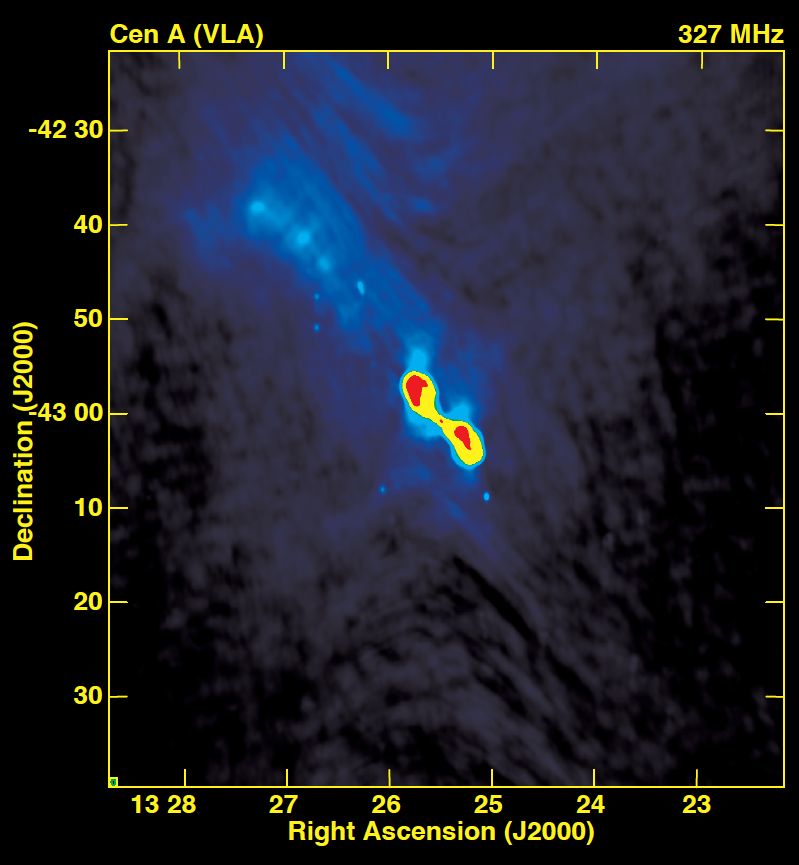}
\caption{Our full-field VLA image at 90 cm/327 MHz, covering the 
NTR and STR regions in addition to bright inner lobes. Diffuse 
emission is apparent throughout the northern half of the image, 
comprising the base of what has been called the North Middle Lobe
(NML).  No comparable emission is seen to the south. 
Three bright radio knots occur along a faint ridge of emission, both 
embedded in a broader region of more diffuse emission.  We see no
indication of a large-scale jet.  Emission is also
detected beyond the previously-known edges of the inner radio lobes (see
Figure {\ref{Fig:InnerLobes}).   Refer to Figure
\ref{Fig:schematic_overall} for details of features we believe to be real 
and/or spurious. The  restoring beam for this 
image is $35\arcsec  \times  25\arcsec  \times 0\arcdeg$.}
\label{Fig:best_overall}}
}
\end{figure*}


We continued with several further cycles of MEM imaging and
self-calibration using the MEM model.  
Final images were produced 
using VTESS.  The four band/pol combinations were each imaged
separately, combined with appropriate weighting
and convolved with a beam just slightly larger than the fits to 
the individual dirty beams (array response functions).
Two images were
produced from the final calibrated data: one of a larger field
including the full region occupied by the North Middle Lobe and any southern
counterpart, the other focusing on the central double source and
inner jet.  The images were corrected for
the  primary beam response of the telescopes.  
Table \ref{Images} lists the properties of the final images used in the paper.

We also produced a spatially filtered image following the method of 
Rudnick (2002), to help in the analysis of small-scale structures 
within the North Middle Lobe, without contamination form the 
diffuse background emission
We used the AIPS task MWFLT to produce an ``open'' filtered image,
which contains all emission from structures larger than $\sim2.7\arcmin$. 
The open filtered image was then subtracted from the original image, 
leaving  only emission from scales smaller than $\sim 2.7\arcmin$ in the 
high-pass filtered image.   
We return to this topic in Section \ref{results_radio_knots}.

\bigskip
\section{Radio structure of  the region  at 90 cm}
\label{full_middle_region}

\begin{figure}[htb]
{\center
\includegraphics[width=.80\columnwidth]{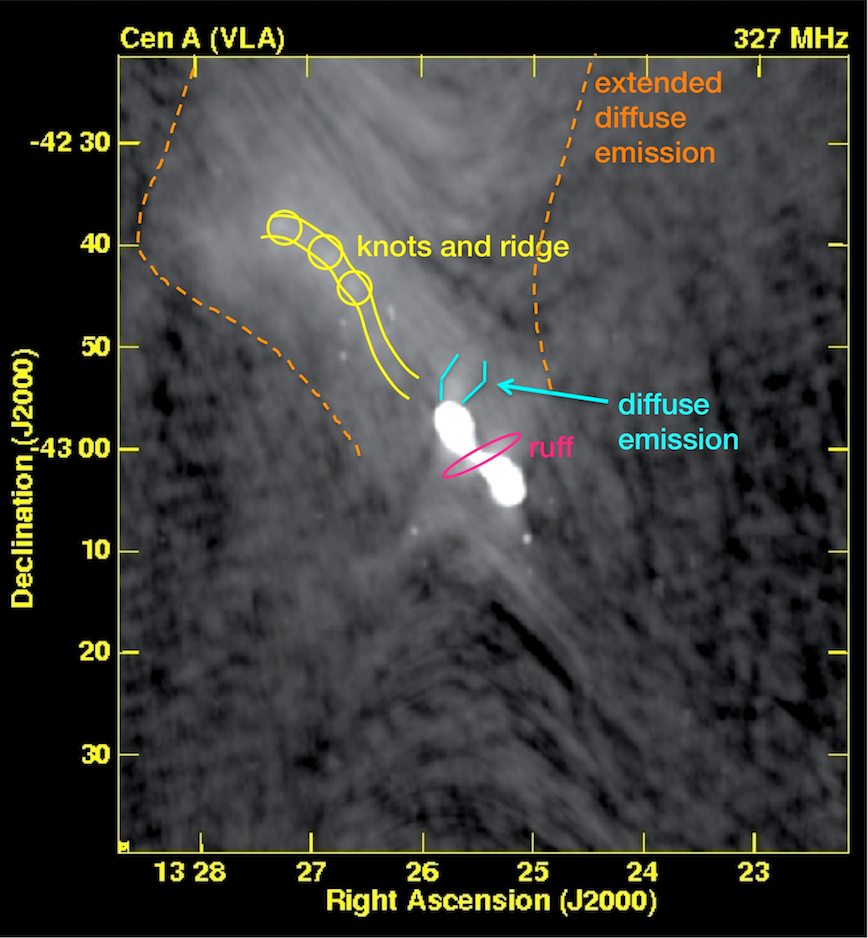}
\includegraphics[width=.93\columnwidth]{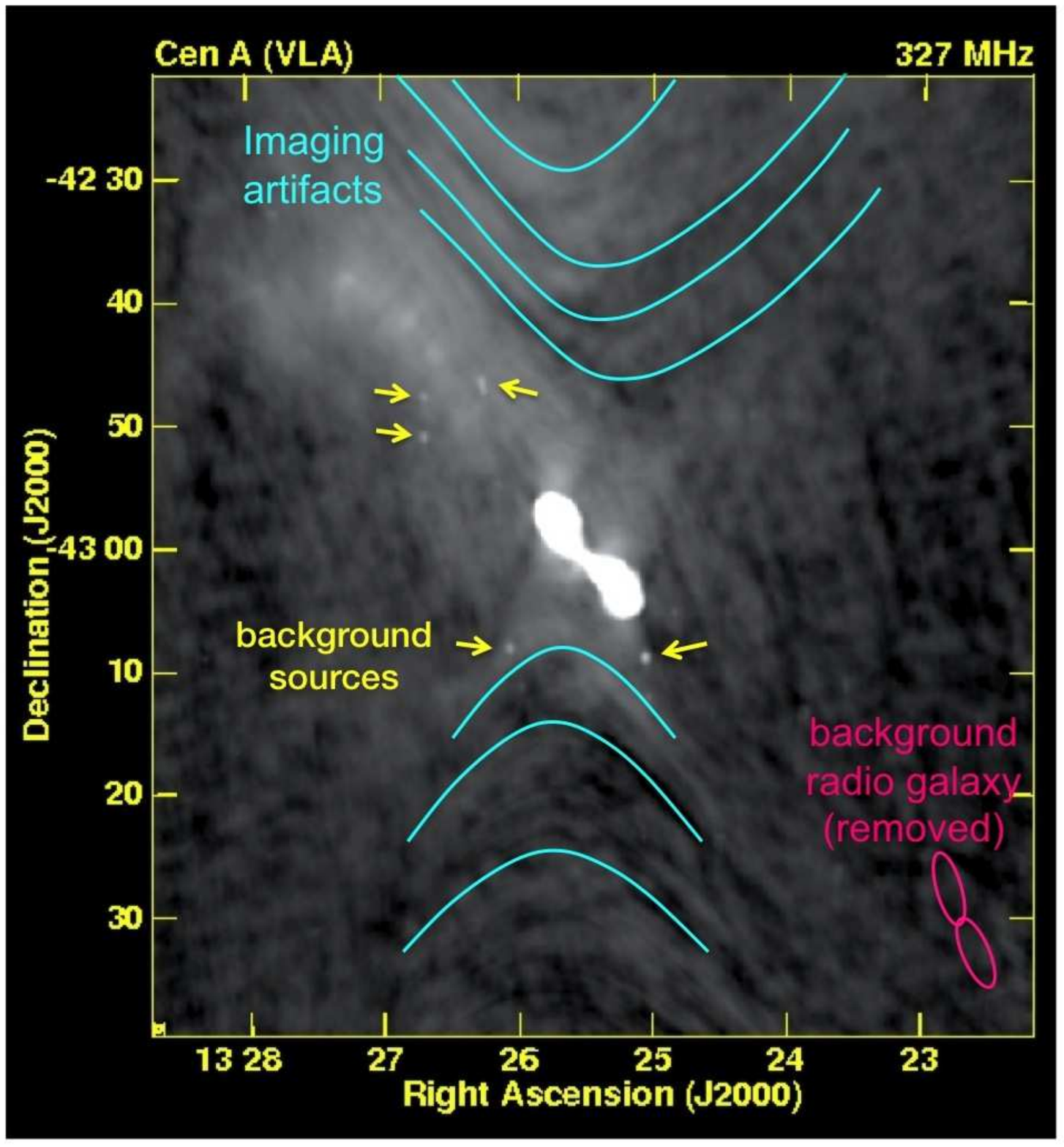}
\caption{Lines overlaid on our full-field, 327-MHz image showing which 
features are where, and indicating their reality (or not).  Upper figure,
what we  believe:  in addition to extended diffuse emission from the
North Middle Lobe, we detect a knotty ridge towards the SE edge of
the North Middle Lobe, and extended emission close to, but outside of,
the Inner Lobes.
Lower figure:  imaging artifacts, as discussed in Section 
\ref{Obs_artifacts}, and background sources. 
  This is the same image as in Figure 2, with a different stretch. 
The  restoring beam  
is $35\arcsec  \times  25\arcsec  \times 0\arcdeg$.}
\label{Fig:schematic_overall}
}
\end{figure}

In Figure \ref{Fig:best_overall}
we show the inner $\sim 90$ kpc of Cen A. The
bright inner lobes are clearly seen, although overexposed.  The 
broad, diffuse emission to the NE is the base of the
North Middle Lobe as seen at 90 cm.   
In the top panel of Figure \ref{Fig:schematic_overall} we present
the same image, but with a deeper stretch, and we label significant
features in our image which we believe are real and which 
are discussed in detail later in the paper.  
These figures reveal key points about the region.

$\bullet$  We detect diffuse emission outside of, but close to,
 the Inner Lobes which has not been reported before.
As indicated in the top panel of 
Figure \ref{Fig:schematic_overall}, 
we see emission extending north from the outer edge of
the North Inner Lobe.  We also detect
extended emission  almost perpendicular to the
Inner Lobes.   We call this latter region ``the ruff'' and
we suggest it is
radio emission from the star-forming disk. We discuss these features
in more detail in Section \ref{results_inner_lobes}.

$\bullet$ In agreement with previous, lower-resolution
 work (Section \ref{TransRegion_overview}), 
we detect  the base of the
 North Middle Lobe as a broad, diffuse structure, 
extending $\sim 10 -  40\arcmin$ 
($\sim 11 -  45$ kpc)  from the North Inner Lobe to the lower region 
of the North Outer Lobe. 
It has a well-defined edge to the SE, in agreement with the 20-cm
image from M99.  The SE edge is roughly coincident with the NW edge of
the X-ray emission from the galactic ISM; as noted above (Section
\ref{TransRegion_overview}), this edge is probably where the thermal
ISM abuts the nonthermal plasma in the North Transition Region.
We discuss this diffuse emission further in Section \ref{results_TRs}.

$\bullet$ We see no evidence for collimated outflow beyond the
North Inner Lobe 
as suggested by M99.  We do, however, see diffuse emission extending
directly north from the North Inner Lobe  which blends into the general
North Middle Lobe.  We also see 
a relatively bright, curved line of emission, which
contains three bright radio knots;  we refer 
to this feature as the ``ridge'' or ``knotty  ridge''.
Both of these features are labelled in the top panel of
Figure \ref{Fig:schematic_overall}.
The south end of the knotty ridge
overlaps the north end of the linear feature 
identified by M99.   This structure lies close to the SE edge of
the North Middle Lobe.  We discuss it further in Section 
\ref{results_radio_knots}, and  we show in Paper 2 that
it  is spatially coincident with the complex 
ribbon of optical emission lines, young stars, Far-UV, 
and X-ray emission which we call the weather system.

$\bullet$  We do not detect any emission from  the South Transition Region, 
down to our sensitivity limit.   As previous authors have
noted, while faint diffuse emission does exist in this region, 
(\eg, Junkes \etal 1993), there is no strong
southern counterpart to the North Middle Lobe.

\subsection{Limitations of our observations}
\label{Obs_limitations}

There are particular challenges to observing a  source so far
south with the VLA.  The source is never more than $12^{\circ}$
 above the horizon; 
 at very low elevations the VLA antennas shadow one another and/or pick up
interfering signals from foreground telescopes; 
loss of coherence due to the ionosphere is exacerbated at low elevations. 
Given the difficulties inherent in these observations, 
we identify several 
effects which may affect image quality, as follows.

\paragraph{Contamination by emission from the outer lobes.} 
All of the structures presented here, including the inner and
transition regions of Cen A, are embedded in 
much larger ($\sim 8\arcdeg$) outer lobes.  Because the outer 
lobes are introduced into
the imaging through the outer parts of primary beam, they can 
appear as peculiar 
structures in the images.   We suspect this is the cause of some of
the noticeable artifacts in our images, which
we were unable to remove.  However, the signatures of these artifacts 
are readily recognized and thus unlikely to be mistaken for source
structures (as discussed in Section \ref{Obs_artifacts}). 

\paragraph{Primary beam effects.} The shape of the 90 cm 
response from VLA antennas is asymmetric and does not 
fall to a full null; rather the sensitivity falls off gradually 
for many degrees. In addition, the
 primary beam is  not circularly symmetric,
which causes structures far from the field center to be modulated as the
parallactic angle changes.  Because the Outer Lobes of Cen A are bright 
at 90 cm,  significant brightness on 
large scales will be detected in the outer parts of the primary beam, which 
cannot be fully removed due to the modulation.  
This large-scale, sidelobe-detected emission, 
collected from an asymmetric beam pattern, 
 appears  in the images as low-level texture in the image background.

\paragraph{Limited uv coverage.} The arrays we
used for these observations are sensitive to structures of size
up to $\sim 1.25\arcdeg$.  The field that was imaged is $\sim 2\arcdeg$,
in each dimension, but the largest images of the North Middle Lobe 
shown in this paper 
are $\sim 1\arcdeg$ across.  Therefore, the structures we show and discuss 
are not affected by missing short uv spacings.  However, we are unable
to image larger-scale emission, because we do not have data 
sensitive to the scales of the outer lobes, as well as the other issues 
with the Outer Lobes discussed above.
Further, the images are affected by missing uv samples at certain angles 
in the outer part of the uv-plane, which  arise because
of the low declination and limited viewing time of the source.


\smallskip

\paragraph{Imaging a large field.} Because the MEM algorithm 
implemented in  AIPS (or CASA) does 
not allow mosaicing of multiple fields on a correct 3D grid, 
we could not account for the curvature of the sky near the edges of the
large image.   
Point sources  will be noticeably distorted and therefore
poorly represented beyond a distance of $\sim 0.3\arcdeg$ from the field
center, and emission on scales $\sim 1\arcmin$ will be affected if it
is further from the field center than  $\sim 0.7\arcdeg$. 
We addressed the possible issue for small
sources by making a
multifaceted image and subtracting the point sources found further 
than $\sim8\arcmin$ from the field center,  as described in section 2.2. 
We avoided problems possibly affecting larger structures by 
imaging fields smaller than the scales that would be affected. 

\subsection{Artifacts in our image}
\label{Obs_artifacts}

In the lower panel of Figure \ref{Fig:schematic_overall}  we indicate
schematically which features in the image we do not believe.
The V-shaped dark gaps swooping across the field, and the parallel
stripes extending diagonally from SW to NE roughly parallel to the
North Middle Lobe's structure, are very likely artifacts of the
image reconstruction, due to the low elevation and resultant
foreshortened visibility of Cen A when viewed from the VLA. 
We also do not believe the dark gap on the NW edge of the North Middle Lobe 
is physical.  Based on previous observations (as mentioned in 
Section \ref{TransRegion_overview}), we
think it likely that the diffuse, North Middle Lobe
emission continues smoothly to the north,
becoming fainter as as it merges into the northern outer lobe and
fades below our detection limits.  The mottling seen in the 
background of Figure \ref{Fig:schematic_overall} probably results
from sidelobe detections of the bright Outer Lobes.

\medskip

\section{Comparison to Morganti (1999) }
\label{compare_to_M99}

\begin{figure}[htb]
{\center
\includegraphics[width=.49\columnwidth, trim=1cm 1cm 1cm 1.8cm, clip=True]{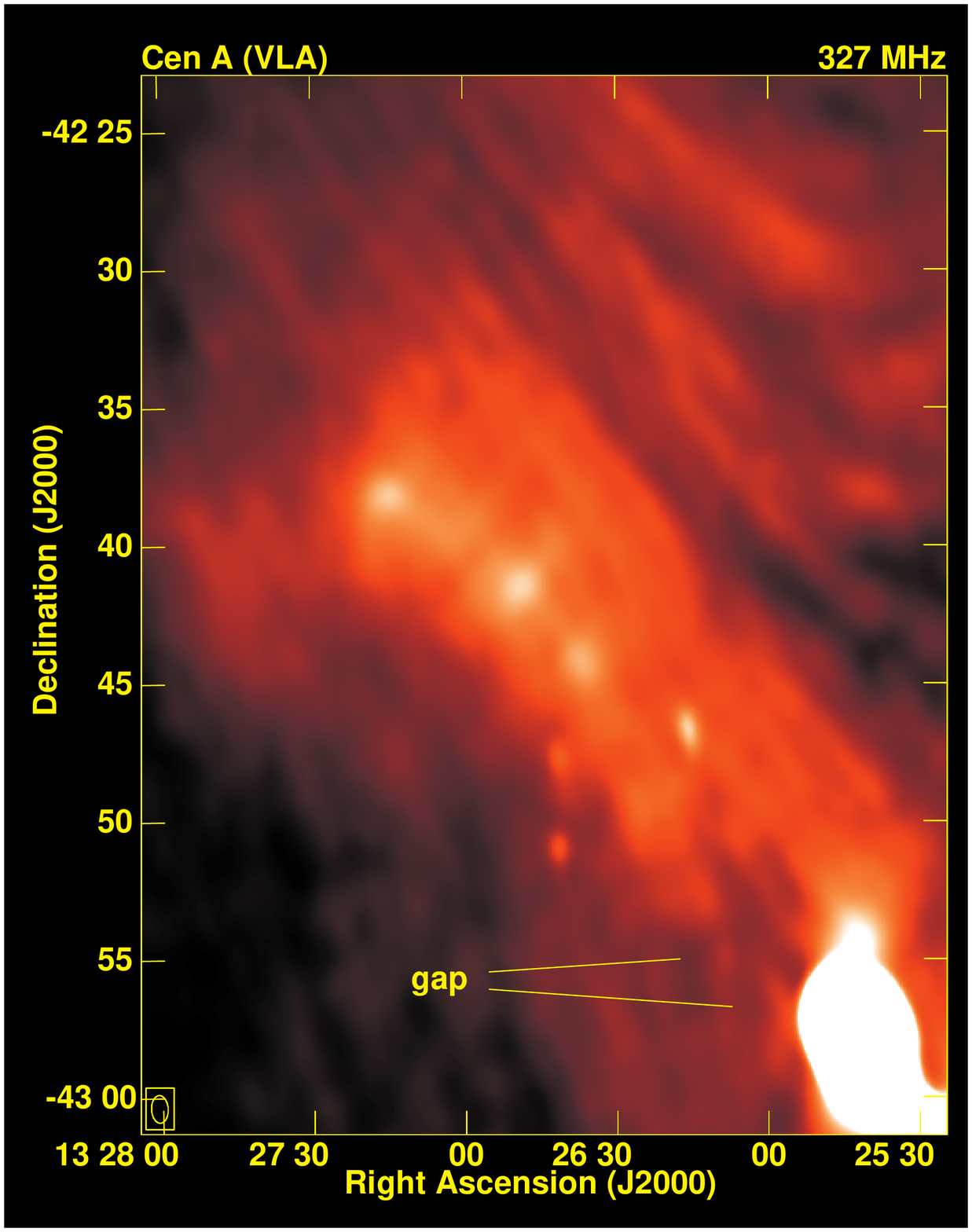}
\includegraphics[width=.485\columnwidth, trim=1cm 1.6cm 1cm 1cm, clip=True]{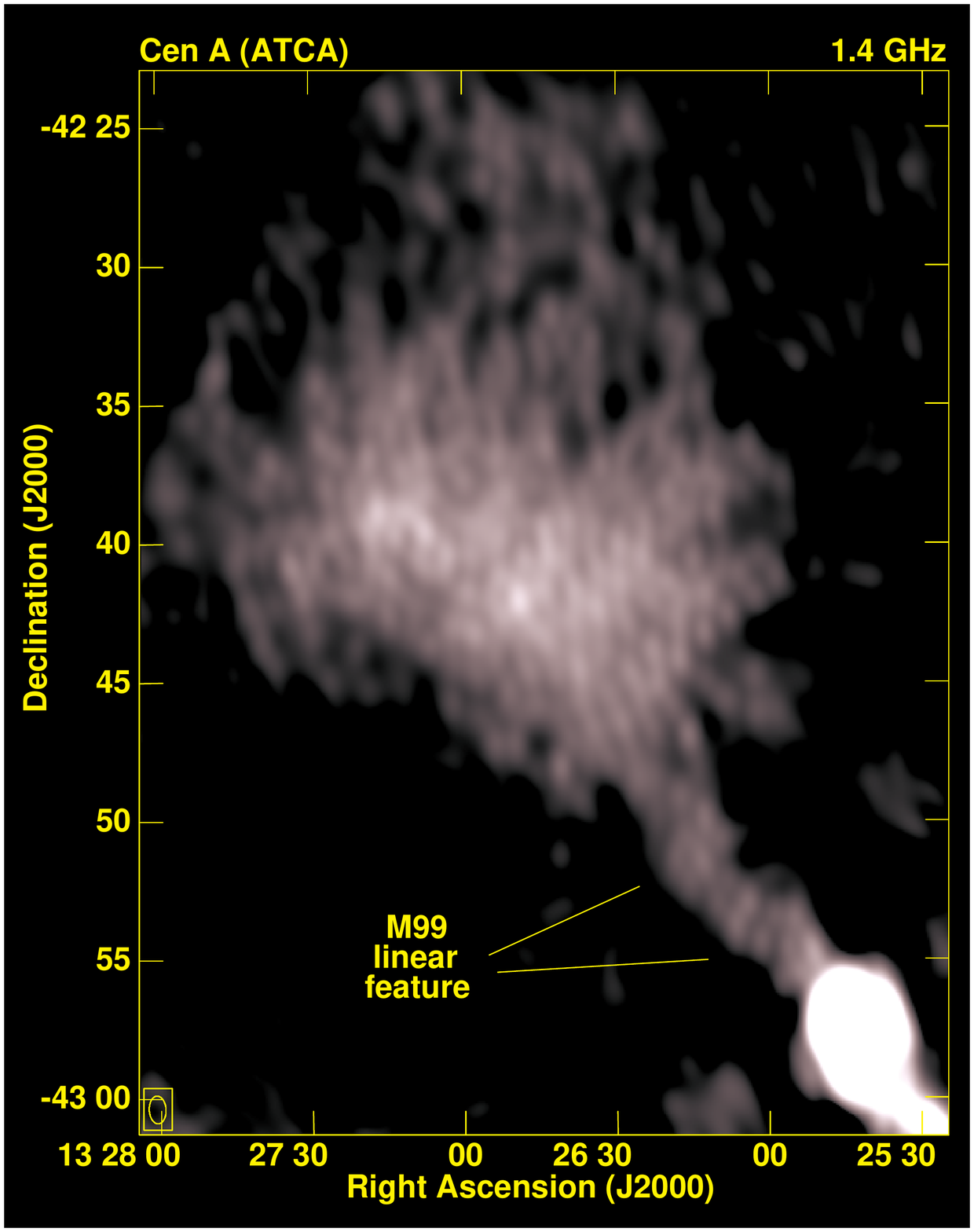}
\caption{ 
The North Middle Lobe, as seen by the VLA at 90 cm (left), and by ATCA at
20 cm (right;  image kindly provided by R. Morganti).  
Both images have a beam
$60.5\arcsec  \times 36.25\arcsec  \times 2.88\arcdeg$.  This comparison
shows that the string of bright knots can be seen at 90 and 20 cm,
as can the   diffuse emission to the NE of the bright ridge.  
However, the linear structure 
connecting the North Inner Lobe to the 
North Middle Lobe in the 20-cm image is not seen at 90 cm;  instead 
we see a gap  just north of the North Inner Lobe,
past  which the radio ridge begins. 
}  
\label{Fig:NML_compare}
}
\end{figure}

Our 90-cm JVLA image of the North Middle Lobe is significantly
different from the 20-cm image from the Australia Telescope Compact
Array (ATCA), presented by M99. In Figure \ref{Fig:NML_compare} 
we show both images, side by side, to highlight the similarities and 
differences.  The images are  convolved to the same resolution and
show identical fields.  The stretch for the 20-cm image was chosen to
approximate the figures in M99 and to avoid showing  unphysical areas of
negative flux.

Although there are broad similarities between the two images, they
disagree in several details. Both images detect extended diffuse emission
from the North Middle Lobe, but that emission is much more extended
in our 90-cm image than it is in the 20-cm ATCA image.  Our 90-cm
image shows no sign of the quasi-linear feature, 
immediately beyond the edge of the North Inner Lobe, which is seen
in the 20-cm image.   Instead, our image actually shows a {\it gap}
-- a region of faint but non-zero emission -- at the same location as the
 inner part of the linear 20-cm feature.
We also see a broad band of emission
extending directly north from the North Inner Lobe and connecting
to the broad NML emission.  Going outwards,
the 90-cm knotty ridge can be seen out to 
$\sim 30\arcmin$ 
 from the nucleus, after which 
it bends to the east and disperses.   By contrast, 
the linear structure in the 20-cm image
is difficult to follow beyond $\sim 15$\arcmin$~$ 
from the nucleus;  past there it disappears into 
the North Middle Lobe. Looking inwards, the ridge in our 90-cm image 
does {\it not} appear to connect the North Middle Lobe to the North Inner Lobe,
but appears instead to bend slightly toward the south and then fade out.  
The north end of the linear M99 feature 
can be seen in our observations, 
but only as a faint enhancement within a much larger region of 
bright, diffuse emission.

\begin{figure}[htb]
{\center
\includegraphics[width=.49\columnwidth, trim=1cm 1cm 1cm 1cm, clip=True]{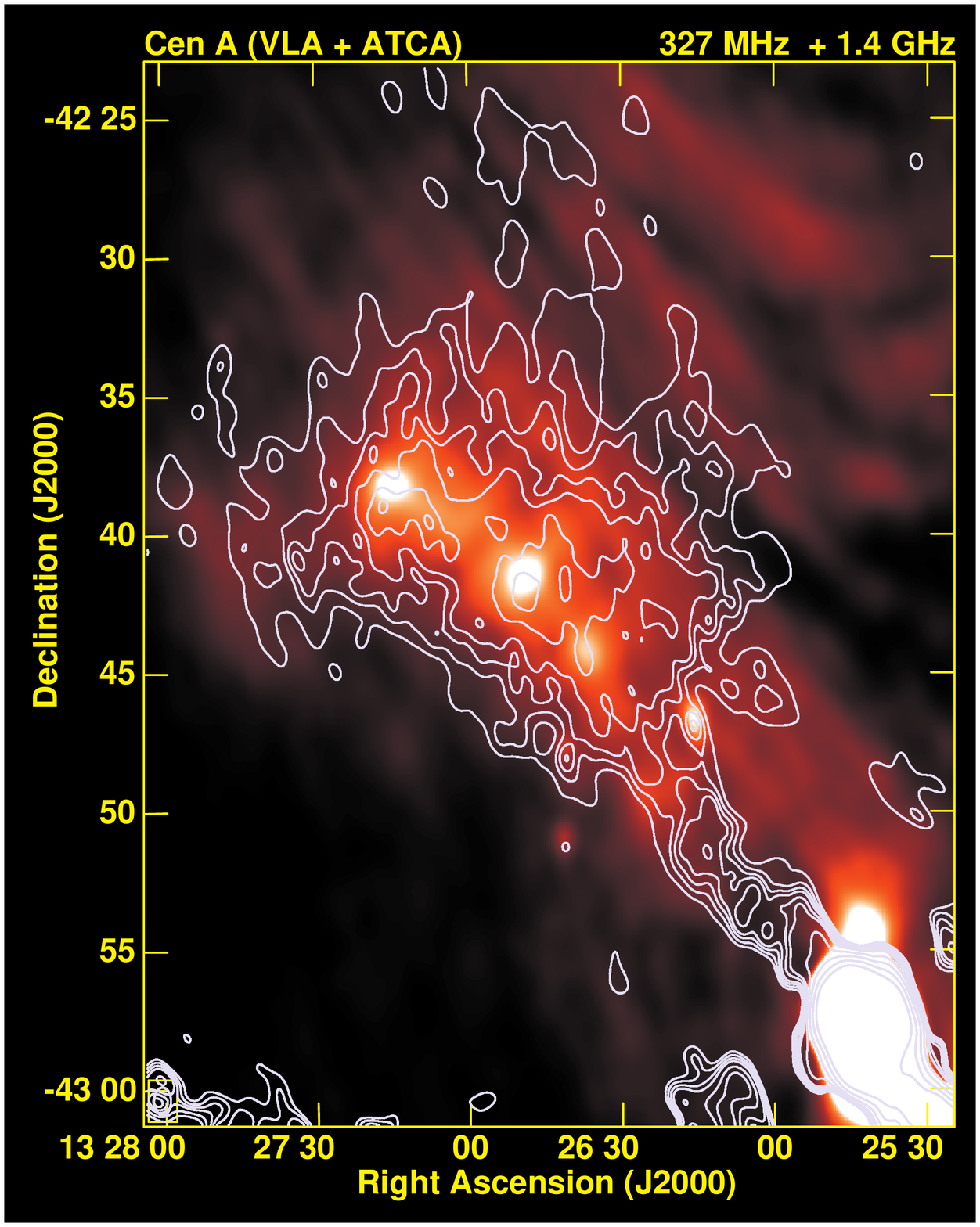}
\includegraphics[width=.48\columnwidth, trim=1cm 1cm 1cm 1cm, clip=True]{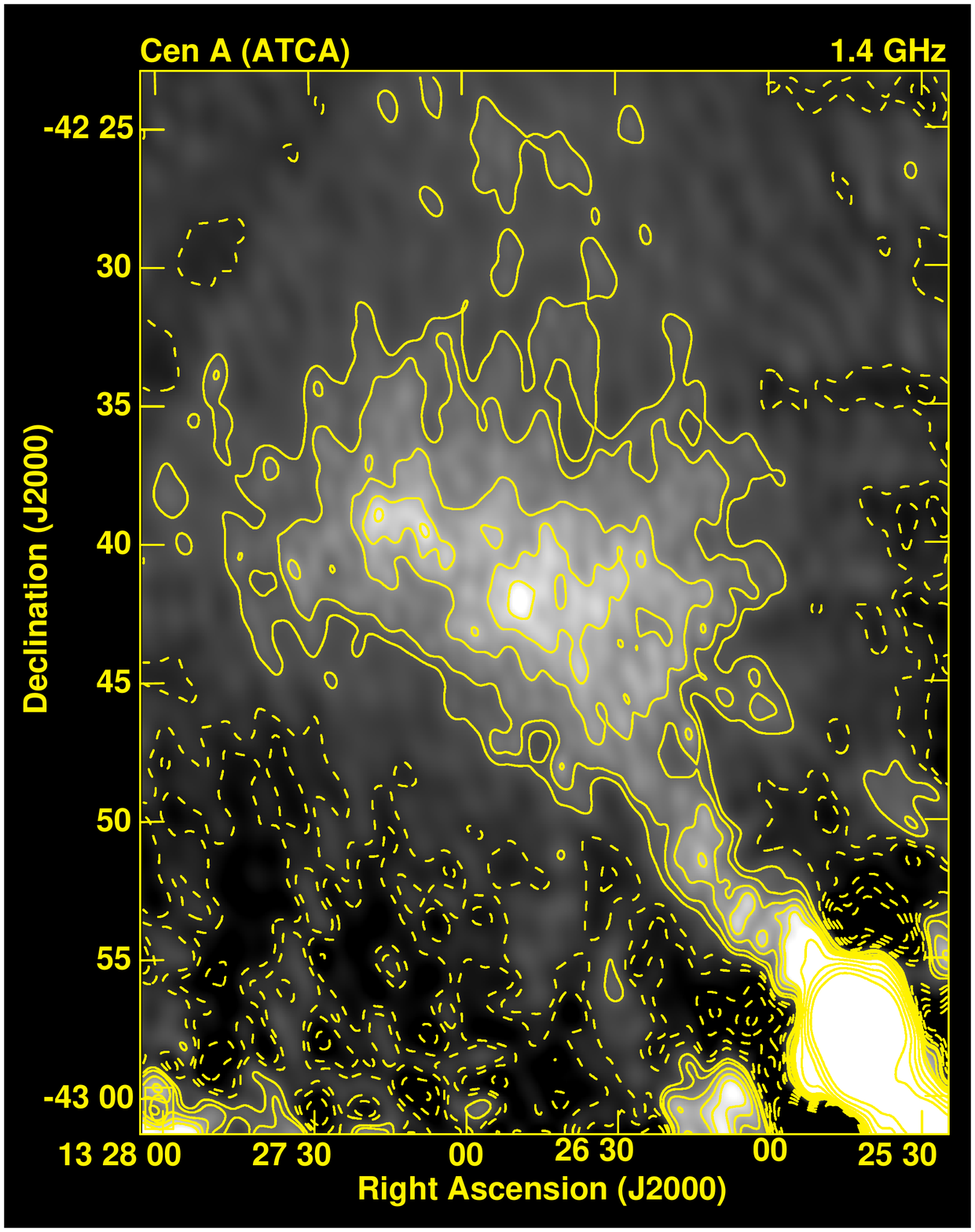}
\caption{Comparing the North Middle Lobe at 20 and 90 cm. 
Left: contours from 20-cm ATCA image (from R. Morganti)
  overlaid on our 90-cm VLA image.  This shows the relation of the linear M99 feature
to the broad band of emission, and the string of bright knots, apparent
in our 90-cm image.  
Contour levels are 1, 2, 3, 4, 6, 8, 16, 32, 64, 128 $\times 12.5$ mJy/beam.
Right:  the 20-cm ATCA image, showing regions of positive and (unphysical)
negative flux. 
Contour levels are -8, -6, -4, -2, -1, 1, 2, 4, 6, 8, 18, 32, 64, 128 $\times
12.5$ mJy/beam;  dotted contours show negative fluxes.
 Both images have the same beam as in Figure  \ref{Fig:NML_compare}.
$60.5\arcsec  \times 36.25\arcsec  \times 2.88\arcdeg$. 
}  
\label{Fig:NML_compare2}
}
\end{figure}

 In Figure  \ref{Fig:NML_compare2} we further compare the two images.
In the left panel of Figure  \ref{Fig:NML_compare2} we show 20-cm contours
overlaid on our 90-cm image.  This shows that 
the linear M99 feature lies to the east of the broad  emission band we
detect at 90 cm, and directly on top of the gap in the
90-cm emission. This figure also shows
that the M99 feature appears to connect directly 
to the knotty ridge we see in the 90-cm image.   In the 
right panel of Figure  \ref{Fig:NML_compare2} we show the 20-cm image with 
stretch and contours chosen to display negative as well as positive fluxes.
This figure reveals  strong negative fluxes in the 20-cm image
on either side of the linear M99 feature. We particularly note a deep negative
hole immediately north of the North Inner Lobe, at the point where we see
a bright 90-cm feature (labelled ``diffuse emission'' in Figure 
\ref{Fig:schematic_overall}) which forms the base of the broad emission
band to the north.

In the rest of this section we discuss possible reasons for the
differences between these two images.

\subsection{Observing and imaging differences}

Differences in the telescope responses of the VLA and the ATCA
are likely to be important in the
 differences between the way diffuse, relatively
faint,  emission is detected in our images and those of M99.
In particular,
accurate detection of faint, diffuse emission in the 
ATCA image may have been hampered by the lower dynamic range of that image.
The central radio source is very bright relative to the North Middle
Lobe at 20cm, 
and ATCA is a relatively sparse array, which can make image reconstruction
challenging, even when the source is not at extreme elevations.

Clearly the deep negative fluxes in ATCA image (right panel of Figure
\ref{Fig:NML_compare2}) are unphysical; we note
the ratio of the brightest part of the linear feature to the most negative
location right next to the linear feature is only $\sim 1.2$.  
These negative patches may be hiding or confusing
detection of diffuse emission in the regions, and make it unclear
how to interpret details of the  structures in the M99 images.

\subsection{Could we see the linear 20-cm feature in our image?}

To test this possibility, we
 measured the brightness of the M99 linear feature
 to be $\sim 37$ mJy/beam at 20 cm. 
If the jet has a spectral index $\alpha = 0.5$ 
($S_{\nu} \propto \nu^{-\alpha}$), it would 
have a brightness at 90 cm of $\sim 77$ mJy/beam; 
if $\alpha = 1.0$ it would have a 90-cm brightness
of 160 mJy/beam. For comparison, the rms in this part of 
our 90-cm image is $\sim 50$ mJy/beam.  Thus, if the M99 
linear feature has $\alpha = 0.5$, it would have  S/N  $\sim 1.5$ 
in our image, \ie, would be essentially undetectable.  If it 
has $\alpha = 1.0$, if would have S/N  $\sim 3$  in our image, 
which would be marginally detectable.
Thus, if the linear M99 feature is a physically real feature, it
must have a relatively flat spectrum in order to not be detected in our
90-cm image.

 We were not able to construct a
meaningful spectral index image due to the differing spatial scales
sampled and to different noise levels in the two images.  
However, by averaging many pixels together, we were able to 
estimate the spectral index in some regions.  Within a region
coincident with the M99 linear feature, we estimate the 
spectal index to be $\alpha \sim 0.7$.
This rough calculation is consistent with our  estimates above:
the linear M99 feature may have a flat enough spectral index that most of its
length is not detectable above the noise in our 90-cm image.\footnote{For completeness, 
we also estimated the mean spectral index in 
the larger, more diffuse regions of the NML which are detected at
both images, finding $\alpha \gtw 1.0$. }  By comparison, 
the Inner Lobes have spectral indices $\alpha = 0.35 - 0.59$ (Clarke \etal 1992b).

\subsection{Missing diffuse flux in ATCA image?}

Another striking difference between our 90-cm image and the 20-cm
ATCA image is the lack of diffuse flux to the west of the 
20-cm linear feature  in the ATCA image.

Could this be because 
the ATCA observations were not sensitive to the diffuse structure?
Our observations included uv spacings both smaller and larger
than those corresponding to the maximum/minimum ATCA baselines used to
produce the image shown in the right panel of Figure \ref{Fig:NML_compare}.
The 90-cm observations are sensitive to emission on
scales up to $\sim 1.25\arcdeg$, whereas the ATCA 
observations were  sensitive to structures $\ltw 25\arcmin$ in size.
Because the 
extended emission we detect from the North Middle Lobe is $\sim
10\arcmin$ across, it should have been visible to the ATCA array.
We checked this by making uniform-weighted images (to match M99) from
our 90-cm VLA data, while restricting the uv ranges to those available
in the two ATCA configurations used in M99. The resulting images still
showed extended emission well beyond that seen in M99.
Thus, we would  have expected ATCA to be sensitive
to  diffuse emission to
the west of the linear feature.

Could the large-scale diffuse
emission be too faint for the  ATCA to detect at 20 cm?  A
typical 90 cm brightness level away from the ridge in the North 
Middle Lobe, sampled
with a beam $135\arcsec \times 75\arcsec$ (the restoring beam in the
ATCA low-resolution image from M99) is $\sim 1$ Jy/beam.  If the extended
source has a typical
spectral index $\alpha \sim 0.7$, 
then the expected brightness in the ATCA
low-resolution image (with rms level $19$
mJy/beam)  would be $\sim 360$ mJy/beam. 
A similar comparison for the high-resolution ATCA image from M99
(with a $56\arcsec \times 36\arcsec$ beam and rms noise 13
mJy/beam), and corresponding flux $\sim 250$
mJy/beam at 90 cm, implies a brightness $\sim 90$ mJy/beam 
expected in the 20 cm ATCA image.  Therefore, we expect the
diffuse emission would have been
 detectable in the ATCA images if it were there at 20 cm,  and
if $\alpha \sim 0.7$. The spectral index of the diffuse emission
would need to be steeper than $\alpha \sim 1.6$ for the 20cm emission 
to be below the detection level of the ATCA image. 

To summarize, we
 suspect that low-level diffuse emission in the south of
the  ATCA image has
been confused by the strong negative  components of the restored ATCA beam. 
  However, it is 
also possible that the diffuse
emission has a much steeper spectrum than the linear M99 feature, or  that
it is extremely smooth and  thus resolved out by the 
ATCA observations.

\subsection{Foreground absorption from the galaxy?}

 Finally, we check whether the absence of the linear M99 feature in our
image could be due to a forergound absorbing cloud.
Such a cloud in the
ISM of NGC 5128  might cause the apparent gap in the 90-cm image, 
where the linear 20-cm  should be seen. 
 The recent detection of an HI cloudlet,
$\sim 11$ kpc from the galaxy (Struve \etal 2010) and spatially
coincident (in projection) with the linear 20-cm feature and 90-cm gap,
suggests that associated ionized hydrogen might be absorbing the
signal.  To check this, we use the expression for free-free
absorption, $\tau_{\nu} \simeq 82. T^{-1.35} \nu_{GHz}^{-2.1} n_e^2
L_{kpc}$ for a cloud of electron density $n_e$ and thickness $L$.
We take the characteristic size of the hole
as $L \sim 10$ kpc.
The most favorable case for detectable absorption is a cloud
at $T \sim 10^4$K.  In order for such a cloud to be opaque at 327 MHz
and transparent at 1.4 GHz, its density must lie in the range $n_e^2
\sim 30 - 600$ cm$^{-6}$ and its pressure in the range $ \sim (1.4\!-\!7.0)
\times 10^{-11}$\dyncm2 ($\sim 16\!-\!80$ times that of the
ISM (Section \ref{results_TRs}) ). 
  Hotter clouds must be at even higher pressures to be opaque at
327 MHz.  It may be that such an over-pressured
cloud just happens to sit in front of
the linear 20 cm feature; but we see no sign of the putative cloud in H$\alpha$
or Far-UV emission (as we show in Paper 2). Thus we judge it unlikely that 
the 90-cm gap is caused by a foreground absorbing cloud.

\subsection{The bottom line: imaging challenges}

Our long list of potential problems, both for our images and
those of M99, mean the final nature of the North Middle Lobe is not 
yet completely clear.  We believe many of the differences between 
our images and those of M99 are  due to a combination of different 
telescope and array responses and the difficulties of imaging faint
diffuse emission in the presence of strong nearby sources.  
The differences between the  images are probably exacerbated 
by spectral index variations across the source. 
The apparent limitations of both images 
highlight the need for improved imaging of this
iconic nearby system. 

In terms of the physics of the region, we agree with previous authors
(including M99)  that the NML is a  region of  broad, diffuse 
emission.  We  caution, however,  that the relatively faint, linear M99
feature is not necessarily a physically meaningful ``jet''.  It
appears to be a relatively bright linear feature
within a fainter, broad emission region, but
detailed interpretion of the  20-cm ATCA image is made difficult 
by  the strong negative patches around the linear feature.  

\bigskip

\section{Radio emission in the inner and transition regions}
\label{Results_section}

In this section we describe and characterize the diffuse emission we
detect in the Inner and Transition regions of Cen A. 
To characterize physical conditions in the  region, we derive the
diagnostic  $p_{min}$: the minimum pressure
the emitting region can have and still produce the observed radio power.  
Although we do not assume $p_{min}$ is the true pressure
of the radio-loud region, comparing $p_{min}$ to the ambient pressure
(for instance known from X-rays) can shed light on the astrophysics 
of the region.

\subsection{Minimum pressures as a diagnostic}
\label{Min_press_text}

We remind the reader of
important uncertanties in the $p_{min}$ calculation. It needs
only two observables -- the flux and area of the emitting region --
but additional assumptions are also required.  One  is the
low-frequency cutoff of the radio spectrum.  The literature 
contains two different choices here.  Some authors (following Burbidge
1965 and Pacholczyk 1970; we call this the ``BP'' method) assume the
radio spectrum extends only to the lowest observable frequencies,
typically 10 MHz. Other authors, motivated by shock acceleration
theory (following Myers \& Spangler 1985; ``MS'' method), assume the
radio spectrum extends to much lower frequencies.  Because the
particle pressure is dominated by the lowest energies in a power law
spectrum, the MS method gives larger $p_{min}$ values than does the BP
method, usually by a factor of a few in the case of Cen A.
  Because both methods have been used for Cen A, we calculate
$p_{min}$ both ways (in Tables \ref{Table:IL_NML_numbers} and
\ref{Table:NML_knot_numbers}).  In our discussion, however, we use
only the more conservative $p_{min}$ from the BP method. 
We relegate the  details of both  methods to 
Appendix \ref{Min_pressure_App}, where the key results are equations
\ref{minP_BP} and \ref{minP_MS}.

Two further uncertainties can be critical to interpreting the
results. The  volume filling factor of relativistic plasma, $\phi$,
is unknown, and can have any value $0 < \phi \le 1$.
Also unknown is 
the ratio of total particle pressure to that in radiating leptons
(the so-called  ``$k$'' factor;  \eg, Burbidge 1959, Pacholczyk 1970). 
We combine both of these uncertainties in
one factor, which we call the {\em pressure scaling factor}
$\eta$, defined in equation \ref{etaBP} for the BP case and
equation \ref{etaMS} for the MS case.  For 
likely conditions in Cen A,  we can approximate both cases
by $\eta \sim  \left[ {( 1 + k) / \phi } \right]^{1/2}$.

What values of $\eta$ might we expect? If
the source contains only radio-loud leptons -- as might, for instance,
describe plasma in the radio jet close to the AGN --  $k = 0$.  If
the same source is also uniformly filled, $\phi = 1$,  then $\eta$ reaches
its minimum, $\eta = 1$. Alternatively,
we know $k \simeq 100$ for galactic cosmic rays, which are dominated by 
relativistic baryons.  Because
the radio-loud plasma we detect in the Transition Regions of Cen A
 is probably drawn from the inner ISM 
of the galaxy, $k \gg 1$ might be more appropriate there.  In addition,
although a uniformly filled volume is possible,
internal filamentation or geometries such as an outer shell of
emission  can give $\phi \ll 1$.  Both $k > 1$ and $\phi < 1$ combine to
give the pressure scale factor $\eta > 1$;  for instance, if $k \sim 100$
and $\phi \ltw 1$ (as suggested by the observed filamentation in
the inner lobes, Clarke \etal 1992b)  we might have $\eta$ on the order
of $\sim 10$. 

In this paper, 
we formally keep $\eta$ as an unknown parameter, but constrain it where
possible from the data.  On larger scales,
we show in  Section \ref{IL_pressures} that  $\eta \sim 10$
is consistent with pressure balance in the Inner Lobes.  In  Section 
\ref{results_TRs} we show that if plasma in the North Middle Lobe
is similar to that in  the Inner Lobes -- if it also has
$\eta \sim 10$  -- then the North Middle Lobe must be at higher 
pressure than its surroundings.

\subsection{The Inner Lobe region}
\label{results_inner_lobes}

\begin{figure}[htb]
{\center
\includegraphics[width=.495\columnwidth, trim=1cm 1cm 1cm 1.5cm, clip=True]{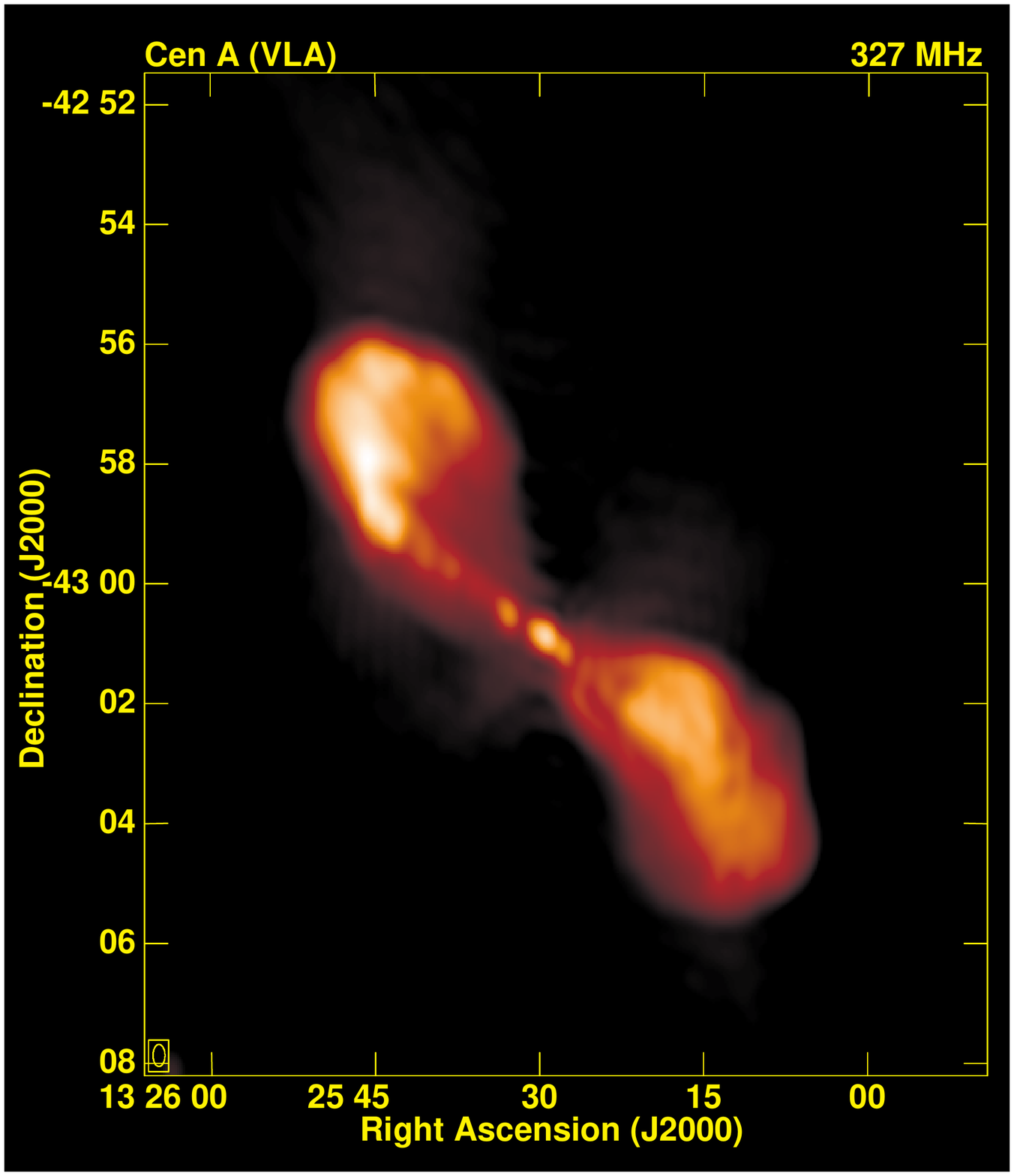}
\includegraphics[width=.495\columnwidth, trim=1cm 1.6cm 1cm 1cm, clip=True]{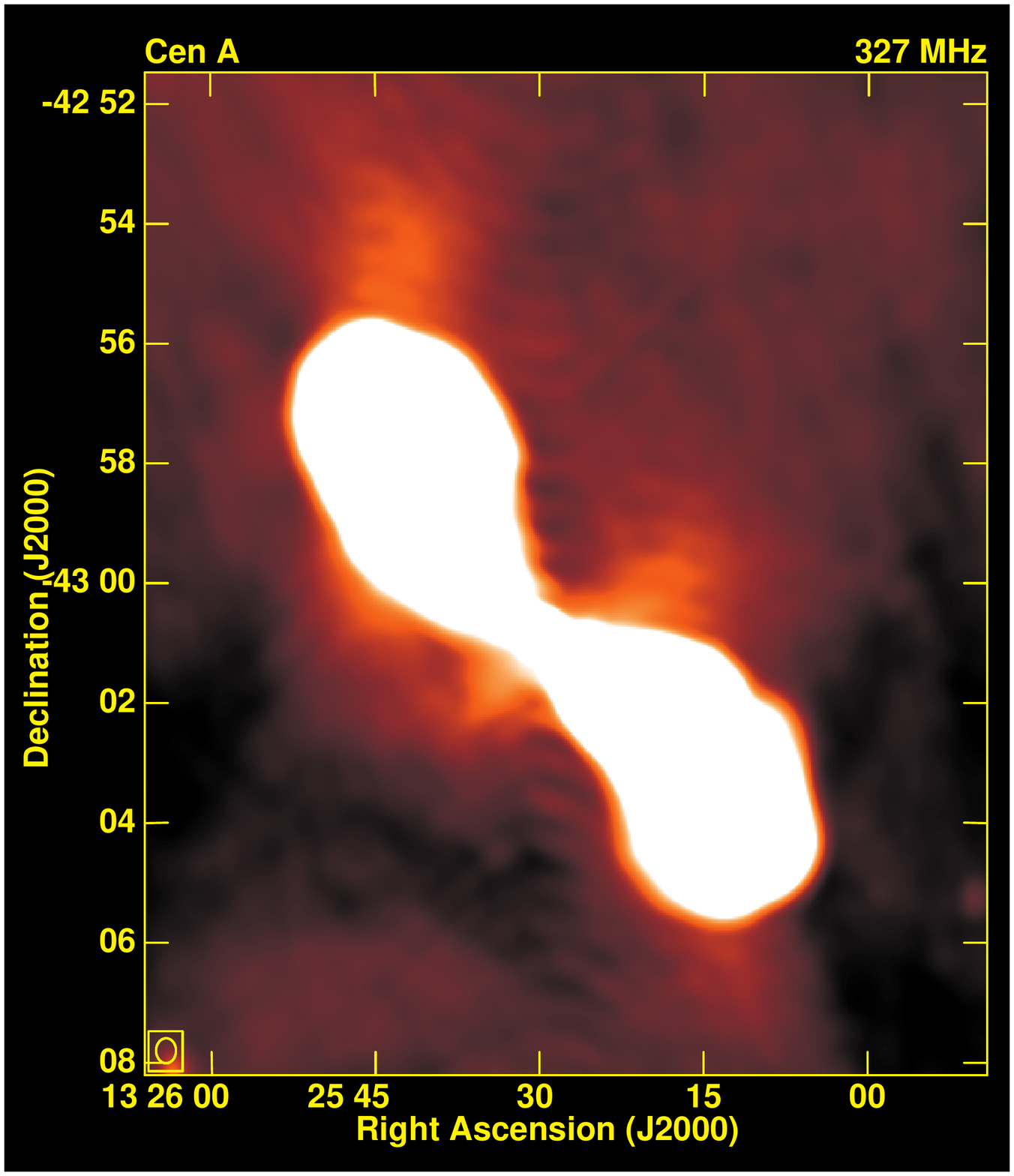}
\caption{The Inner Lobes as seen at 327 MHz.  The left panel
 shows the previously 
known NIL and SIL;  structure is very similar to that seen at higher freqencies 
(1.5 and 5GHz, see text) and here.  The image is 
slightly convolved to $22\arcsec  \times  12\arcsec \times 0\arcdeg$.
The right panel is a highly stretched image, convolved to $25\arcsec 
\times  20\arcsec \times 0\arcdeg$ 
to show the diffuse emission detected beyond the previously known boundaries
of the inner lobes.   We believe the
 bright ``ruff'', roughly perpendicular to the
inner jet, is related to star formation in the galaxy disk. 
Extended diffuse emission is also seen to the N, and SE of the North 
Inner Lobe, and to the S of the South Inner Lobe. 
\label{Fig:InnerLobes}}
}
\end{figure}

Figure \ref{Fig:InnerLobes}a
 shows our high resolution, 90 cm image of the inner $\sim 15$ kpc of Cen A,
displayed to show the well-known bright features in the North Inner Lobe
and South Inner Lobe.
Figure \ref{Fig:InnerLobes}b shows the same field, but with the 
display stretched to emphasize diffuse emission we detect past the
sharp edges of the Inner Lobes.

\subsubsection{Inner lobe morphology}
The morphology of the Inner Lobes in our image
is very similar to that seen at 18 cm and 6
cm (Burns \etal 1983, Clarke \etal 1992b). Both Inner Lobes have sharp
boundaries, and are much brighter than the more diffuse emission
farther north in the North Middle Lobe.  There is no 
indication of a violent "breakout" around the North 
Inner Lobe at the location of the base of the linear M99 feature 
(as might be expected if that feature were a jet extending 
beyond the Inner Lobe).  The
northern inner jet is clearly visible, extending out from the AGN,
flaring out and decollimating at $\sim 4.2\arcmin (\sim 5$ kpc) from the
core.  We confirm previous results in not detecting any southern
kpc-scale jet within the South Inner Lobe.


 \begin{table*}[htb]
\caption{ Synchrotron Analysis for Inner Lobes and North Transition Region}
\label{Table:IL_NML_numbers}
\begin{center}
\begin{tabular}{l l l l l l l c c }
\hline
\rule{0pt}{16pt} \hspace{-8pt} 
Region & Flux  &  Area    & geometry & distance$^a$  & volume &  Figure  
 &$p_{min}^{BP}$ & $p_{min}^{MS}$
\\
    & (Jy)     &(amin$^2$) & &  (amin) & (amin$^3$) &  
&    (dyn/cm$^2$)$^b$  & (dyn/cm$^2$)$^b$
\\[4pt]
\hline 
\rule{0pt}{16pt} \hspace{-8pt} 
Inner lobes:
\\[4pt]

$\quad$ entire NIL    & 380  & 15.1& ellipsoid& $\sim$3 & 34.3 & \ref{Fig:green_blotches_ILs}a
& $1.3\eta\times 10^{-11}$ &  $ 2.9\eta\times 10^{-11}$ 
\\[4pt]
$\quad$ NIL, outer aret   & 338  & 10.1 & sphere&  $\sim$4.5  & 24.3 & \ref{Fig:green_blotches_ILs}b
 & $1.5\eta\times 10^{-11}$ &   $3.3\eta\times 10^{-11}$ 
\\[4pt]
$\quad$ entire SIL   & 282  & 15.8  & ellipsoid &  $\sim$3  &38.8 & \ref{Fig:green_blotches_ILs}c
      & $1.0\eta\times 10^{-11}$ & $2.4\eta\times 10^{-11}$ 
\\[4pt]
$\quad$ SIL, inner part  & 97  & 2.41  & sphere & $\sim$2 & 2.81 & \ref{Fig:green_blotches_ILs}d 
      & $2.5\eta\times 10^{-11}$ &  $5.4\eta\times 10^{-11}$ 
\\[4pt]
$\quad$ SIL, outer part  & 94  & 3.93 & sphere & $\sim$ 4.5  &  5.87 & \ref{Fig:green_blotches_ILs}e
      & $1.6\eta\times 10^{-11}$& $3.6\eta\times 10^{-11}$ 
\\[4pt]
\hline
\rule{0pt}{16pt} \hspace{-8pt} 
Near, but outside Inner Lobess:
\\[4pt]
$\quad$ local region \#1   & 2.73  & 28.8    & pencil beam $^c$& $\ltw 6$ 
&$ 288$ & \ref{Fig:green_blotches_MLs}f    & $ 8.5\eta\times 10^{-13}$ &  $ 2.2\eta\times 10^{-12}$ 
\\[4pt]
$\quad$ local local region \#2  & 5.21  & 32.5   & pencil beam$^c$& $\gtw 8$
  &$ 325$ & \ref{Fig:green_blotches_MLs}g    & $1.2\eta\times 10^{-12}$ & $3.0\eta\times 10^{-12}$ 
\\[4pt]
\hline 
\rule{0pt}{16pt} \hspace{-8pt} 
The NML:
\\[4pt]
$\quad$ 
Kraft comparison, cylinder  & 112  & 350 &   cylinder& $\sim$20  & 3670  & \ref{Fig:green_blotches_MLs}a 
  & $4.5\eta\times 10^{-13}$ & $1.2\eta\times 10^{-12}$ 
\\[4pt]
$\quad$ Kraft comparison, pencil  & 112  & 350 &   pencil beam$^c$
& $\sim$20  & 5250 & \ref{Fig:green_blotches_MLs}a 
  & $3.6\eta\times 10^{-13}$ &  $1.0\eta\times 10^{-12}$ 
\\[4pt]
$\quad$ full NML, ellipsoid  & 394  & 2060 &  ellipsoid 
& $\sim$35 & 61,700 &  \ref{Fig:green_blotches_MLs}c
& $1.8\eta\times 10^{-13}$ &  $3.5\eta\times 10^{-13}$ 
\\[4pt]
$\quad$ full NML, pencil  & 394  & 2060 &   pencil beam$^c$
& $\sim$35 & 32,900&  \ref{Fig:green_blotches_MLs}c
& $ 2.7\eta\times 10^{-13}$ &  $7.5\eta\times 10^{-13}$ 
\\[4pt]
$\quad$ Diffuse, outer, west  & 20.2  & 62.6 &   pencil beam$^c$
& $\sim$23 & 941 & \ref{Fig:green_blotches_MLs}d
  & $3.7\eta\times 10^{-13}$ &  $1.0\eta\times 10^{-12}$ 
\\[4pt]
$\quad$ Diffuse, inner, saddle  & 56.9  & 220 &  pencil beam$^c$ 
& $\sim$12  & 3290  & \ref{Fig:green_blotches_MLs}e
  & $3.2\eta\times 10^{-13}$ &  $ 9.0\eta\times 10^{-13}$ 
\\[4pt]
\hline
\end{tabular}
\end{center}
$^a$ Distances are from center of
galaxy (radio core) to center of region measured.  
\\
$^b$ 
See Section \ref{Min_press_text} for  definition of the pressure
scale factor $\eta$.  We suggest that $\eta \sim 10$ in the Inner 
Lobes and in the diffuse North Middle Lobe (Sections 
\ref{IL_pressures} and \ref{Diffuse_within_NML}).
See equations (\ref{useful_relations_BP},
\ref{useful_relations_MS}) to convert $p_{min}$ to $B_{min~p}$.
\\
$^c$
Pencil beam depth, $L$,  scaled to $10\arcmin$ for region around inner lobes,
(excluding featues identified in Figure \ref{Fig:schematic_overall})
and to $15\arcmin$ for diffuse emission in transition  regions.  
(See Section \ref{Diffuse_around_ILs}
Other choices for line of sight depth, $L$ give 
$p_{min}^{BP} \propto 1/L^{.581}$ and $p_{min}^{MS} \propto 1/L^{.540}$.
\\[10pt]
\end{table*}


\subsubsection{Inner lobe pressures}
\label{IL_pressures}

We measured radio
fluxes in several areas of our image, using the AIPS program BLSUM
to select regions within which the flux density was summed.
Our results are in Table \ref{Table:IL_NML_numbers};  the selected
regions are shown in Figure \ref{Fig:green_blotches_ILs} (online material).
 To get average values through each
inner lobe, we measured the flux within the entire lobe. 
We treat each lobe as a prolate ellipsoid lying in
the sky plane.  We also measured fluxes within smaller, brighter areas
within each lobe, assuming a spherical geometry to derive the sampled
volume.  

We find  
$p_{min} \sim \eta \times 10^{-11} $ \dyncm2 in the outer parts of
both Inner Lobes, and a bit larger for inner part of the South Inenr 
Lobe.
To compare $p_{min}$ to {\green the} pressure of the local ISM, we turn to X-ray data. 
Croston \etal (2009) estimate $p_{ISM} \sim 1.3 \times 10^{-12}$\dyncm2 for 
the ISM  $ \sim 6'$ south of the core (ahead of the X-ray shock
which bounds the South Inner Lobe).  Kraft \etal (2008)
estimate $p_{ISM}\sim 3.0 \times 10^{-12}$\dyncm2 for the ISM
 $\gtw 3.2'$ north of the core. We therefore agree with previous
authors (Burns \etal 1983, Clarke \etal 1992b), that
minimum pressure in the Inner Lobes is well
above that of the ISM, so that the Inner Lobes should be 
expanding into the ISM. 

These results allow us to constrain the pressure factor $\eta$ in the
Inner Lobes. If the South Inner Lobe is driving a shock ahead 
of itself into the
ISM, its pressure should be comparable to that in the shock.  Croston
\etal (2009) find $p_{shock} \sim 1.1 \times 10^{-10}$\dyncm2; in
order to have $p_{min}^{SIL} \ltw p_{shock}$, we require $ \eta \ltw
10$ in the South Inner Lobe.  This value is roughly consistent 
with our suggestions
in Section \ref{Min_press_text} that $k \sim 100$, $\phi \ltw 1$ in
and beyond the Inner Lobe region.

\subsubsection{Extended emission around the Inner Lobes}
We detect faint but real radio emission beyond the apparent Inner Lobe
boundaries. We have stretched the display in the right panel of Figure
\ref{Fig:InnerLobes} to show this clearly, and have indicated the
features schematically in the top panel of Figure
\ref{Fig:schematic_overall}.  We see broad, diffuse emission beyond
the north edge of the North Inner Lobe, extending into the North
Middle Lobe region, but not at the right position angle or
sufficiently collimated to be the base of M99's suggested outer jet.
We also see a ``ruff'' of diffuse emission on either side of the
``waist'' of the inner lobes, with an extent $\sim 5.3$\arcmin
  ($\sim 6$ kpc) from NW to SE of the nucleus.  This ``ruff'' is
  coincident with the dust lane and appears to be radio
  emission coming from the star-forming disk. We discuss this
  further in paper 2.  More diffuse emission appears to extend from
this ruff and wrap around to the NE, around the edge of the North
Inner Lobe.  Finally, we also see diffuse emission extending beyond
the south end of the South Inner Lobe.

\subsection{Diffuse emission in the Transition Regions}
\label{results_TRs}

In this section we characterize diffuse emission in three regions:
within the large-scale North Middle Lobe, close to but outside of 
the Inner Lobes, and at the location where one might expect a South
Middle Lobe. 
Our measurements and derived pressure estimates are
collected in Table \ref{Table:IL_NML_numbers} and 
summarized in Table \ref{Table:pressures}.

\subsubsection{Diffuse emission within the North Middle Lobe.} 
\label{Diffuse_within_NML}

Figure \ref{Fig:best_overall}
shows that the North Middle Lobe seen at 90 cm is a broad, diffuse structure.
We see no sign of a collimated jet, but we do see
three bright compact knots, close to but inside the SE edge of 
the NML (explored further in Section \ref{results_radio_knots}).
While the dark diagonal bands cutting through the North Middle Lobe are
probably imaging artifacts (Figure \ref{Fig:schematic_overall}), 
the lack of emission
to  the SE of the North Middle Lobe agrees with previous authors ({\it e.g.,} 
M99) and is likely real.  This SE edge is where 
the radio-loud North Middle Lobe plasma abuts the thermal ISM of the galaxy. 

To characterize the diffuse emission in the North Middle Lobe,
we measured the flux within several hand-drawn regions.  One
was a rectangular box, of $6.5' \times 21.7'$, meant to reproduce the area
measured by K09.   We also measured two
 other irregular regions of comparable area.
 Finally, we measured the total flux
from a larger area, meant to include the full North Transition Region,
 including lower surface
brightness areas. All areas are shown in Figure \ref{Fig:green_blotches_MLs},
and our results are given in Table \ref{Table:IL_NML_numbers}.
To convert to volumes we either assumed ellipsoidal or cylindrical
geometry in the sky plane,  or a pencil beam, $15\arcmin$  
through the source, as appropriate for each measurement.  
The numerical range seen in Table \ref{Table:IL_NML_numbers}
comes from differences among the measured regions,
different assumed geometries, and different assumptions in the 
minimum-pressure calculations (as in Section \ref{Min_press_text}). 
In Table \ref{Table:pressures}
we extract from this range a ``characteristic'' $p_{min}$ for the North
Middle Lobe, $p_{min} \gtw 4 \eta \times 10^{-13}$\dyncm2.

 \begin{table*}[htb]
\caption{ Various pressures in the North Transition Region}
\label{Table:pressures}
\begin{center}
\begin{tabular}{l c c l}
\hline
\rule{0pt}{12pt} \hspace{-8pt} 
Region/object & pressure & reference & comments
\\[2pt]
& (\dyncm2) & 
\\[2pt]
\hline
\rule{0pt}{12pt} \hspace{-12pt} 
Around the ILs:
\\[2pt]
thermal  ISM (X-rays) & $\sim 1-3\times 10^{-12}$  & Section \ref{Min_press_text}
   &   $\sim 3-6\arcmin$ from core
\\[2pt]
inner diffuse radio & $\gtw \eta \times 10^{-12}$ & Table \ref{Table:IL_NML_numbers} 
  &  $p_{min}$ from synchrotron
\\[2pt]
\hline
\rule{0pt}{12pt} \hspace{-12pt} 
Within the NML:
\\[2pt]
thermal  ISM (X-rays) & $\sim 9 \times 10^{-13}$ 
 & Section \ref{Diffuse_within_NML}  &  $\ltw 30\arcmin$ to SE
\\[2pt]
diffuse NML (radio) & $\gtw 4 \eta \times 10^{-13}$
& Table \ref{Table:IL_NML_numbers} 
&  $p_{min}$  from synchrotron
\\[2pt]
\hline
\rule{0pt}{16pt} \hspace{-12pt} 
Compact features:
\\[2pt]
X-ray knots (if thermal)  & $\sim 2 \times 10^{-11}$  &
Section \ref{results_radio_knots} & 
see Section \ref{NML_Xrayknots} for alternatives
\\[2pt]
radio knots (if synchrotron)  & $\gtw \eta \times 10^{-12}$& 
Table \ref{Table:NML_knot_numbers} & $p_{min}$
from synchrotron
\\[2pt]
\hline
\end{tabular}
\end{center}
Synchrotron-based $p_{min}$ values use the more conservative BP
method; see Appendix \ref{Min_pressure_App}. 
The  pressure scaling factor $\eta $ (Section \ref{Min_press_text}) 
may vary  with location;  
we suggest that $\eta \sim 10$ in the Inner 
Lobes and in the diffuse North Middle Lobe (Sections 
\ref{IL_pressures} and \ref{Diffuse_within_NML}), and that 
$\eta \sim 10$ is a reasonable assumption for the radio knots
(Section \ref{NML_Xrayknots}).
\\[10pt]
\end{table*}

To compare this to the local ISM, we need the pressure of the ISM and
the likely value of the pressure scale factor $\eta$.  For the ISM, we
use results from K09, who found $p_{ISM} \sim 8.5 \times
10^{-13}$\dyncm2 for a wedge of the X-ray halo SE of the radio source
out to $\ltw 30'$. We assume this is the dominant pressure in the
region.  For the pressure scale factor $\eta$, we  noted in Section
\ref{Min_press_text} that $\eta$ can be as small as unity.  This
holds if the  radio-loud plasma is
homogeneous (filling factor of order unity) and if cosmic-ray baryons
provide no more pressure than the radio-loud electrons (unlike the 
galactic ISM).  If $\eta \sim 1$, and if $p_{min}$ is close to
the true pressure within the North Middle Lobe (meaning that
 neither the electrons nor the magnetic  field greatly
dominate the pressure), then the
diffuse North Middle Lobe can be in pressure balance with its surroundings.

However, there may be a good case for expecting a larger value of
$\eta$ in the region.  We pointed out in Section \ref{Min_press_text}
that $\eta \sim 10$ would be expected if the radio-loud plasma
contains cosmic ray baryons comparable to those in the ISM (where
baryon pressure $\sim 100$ times larger than lepton pressure).
Furthermore, in Section \ref{IL_pressures} we showed that $\eta \sim
10$ is consistent with pressure balance between the South Inner Lobe
and the X-ray-loud shock it is driving into the surrounding ISM.  If
this larger $\eta \sim 10$ also applies to the radio-loud plasma in
the diffuse North Middle Lobe, then the North Middle Lobe  
is at a substantially higher pressure than the local ISM.

\subsubsection{Diffuse emission close to the Inner Lobes.}
\label{Diffuse_around_ILs}
To describe the diffuse emission close to, but outside of, the Inner
Lobes, we
measured fluxes in two areas, $\sim 6\arcmin$ and $\sim 8\arcmin$ away
from the galaxy (shown in online material in 
Figure \ref{Fig:green_blotches_MLs}).
We treated each as a pencil beam, taking the
depth along line of sight to be $10' (\sim 11 $ kpc).  We find
$p_{min} \sim \eta \times 10^{-12} $\dyncm2 for the two regions.
Again comparing to ambient ISM pressure (from
Table \ref{Table:pressures}), we see that -- similarly to the situation
in the broader diffuse emission throughout the North Middle Lobe -- 
the plasma in  this region around the Inner Lobes 
is at higher
pressure than the local ISM, if the scale factor $\eta \sim 10$.

\subsubsection{Non-detection of a South Middle Lobe} 
We do not detect any emission from the region where we might expect a 
South Middle Lobe (SML) to be.
To determine an upper limit for the radio luminosity of a 
South Middle Lobe, we measured fluxes in two boxes of the 
same dimensions as those used for the 
North Middle Lobe.
One box was at the location where we would expect to see a 
South Middle Lobe, and another
well away from that location (shown in Figure \ref{Fig:green_blotches_MLs}).
In the ``off-source'' regions, we measured a $1 \sigma$ noise level of $2.6$
mJy/beam with a $35\arcsec \times 25\arcsec \times 11\arcdeg$ beam, 
or 6.1 Jy within the box.  We set the detection limit for
any SML flux within the North Middle Lobe-equivalent box at 
$3 \sigma \sim 18.3$ Jy. This implies a NML/SML brightness ratio  of at 
least 5-6, in
agreement with the observed NML/SML surface brightness ratio
$\sim 10$ in the Junkes \etal (1993) 4.8 GHz (6.3 cm)  image, which had a
$4.3\arcmin \times 4.3\arcmin$ beam.

\bigskip

\section{Compact features in the North Middle Lobe}
\label{results_radio_knots}

While we see no evidence of a collimated jet past the north edge of
the North Inner Lobe, we do detect three compact knots of emission,
close to the SE edge of the North Middle Lobe and  apparently
connected by a fainter, narrow ridge of emission. 
These features 
are highlighted in a spatially filtered image shown
in Figure \ref{Fig:radio_knots}.  The ridge extends $13\arcmin 
- 30\arcmin$ ($\sim 15 - 34$ kpc) from the galactic core, and the 
knots occur at distances $21\arcmin - 30\arcmin$ ($\sim 24 - 30$ kpc) 
from the core.  The ridge appears to overlay part of, and to be an
extension of, the linear feature reported at 20 cm by M99, and the radio-loud
knots  are near local maxima in the M99 image (Figures
\ref{Fig:NML_compare}, \ref{Fig:NML_compare2}).

\subsection{Measurements: radio knots}
\label{measurements_radio_knots}

\begin{table*}[htb]
\caption{North Middle Lobe Radio Knot Properties: 
} 
\label{Table:NML_knot_numbers}
\begin{center}
\begin{tabular}{c c c c c c c c}
\hline
\rule{0pt}{16pt} \hspace{-8pt} 
Region & RA  & Dec  & Flux  &\multicolumn{2}{c}{----Radius$^a$----} 
&   $p_{min}^{BP}$  & $p_{min}^{MS}$ 
\\[4pt]
  & &  & (Jy)   & (arcsec)  & (pc)      
        &  (dyn/cm$^2$)$^b$   &(dyn/cm$^2$)$^b$ 
\\[4pt]
\hline 
\rule{0pt}{16pt} \hspace{-8pt} 
North knot & 13$^{\rm h}$27$^{\rm m}$14.6$^{\rm s}$ & -42\deg38\arcmin07\arcsec   
& 0.164  &  36  & 660 
   & $1.2\eta \times 10^{-12}$ & $3.2\eta \times 10^{-12}$ 
\\[4pt]
Middle knot& 13$^{\rm h}$26$^{\rm m}$37.2$^{\rm s}$ & -42\deg44\arcmin05\arcsec    
& 0.210   &  42  & 770 
    &$1.1\eta \times 10^{-12}$ & $2.9\eta \times 10^{-12}$ 
\\[4pt]
South knot & 13$^{\rm h}$26$^{\rm m}$48.6$^{\rm s}$ & -42\deg41\arcmin26\arcsec  
& 0.136   &  35  & 640 
& $1.2\eta \times 10^{-12}$ & $3.0\eta \times 10^{-12}$ 
\\[4pt]
\hline
\end{tabular}
\end{center}
$^a$ Spherical geometry assumed for each knot; 
 radii determined as discussed in text.
\\
$^b$ See Section \ref{Min_press_text} for  definition of pressure 
scale factor $\eta$. We suggest that $\eta \sim 10$ in the Inner 
Lobes and in the diffuse North Middle Lobe (Sections 
\ref{IL_pressures} and \ref{Diffuse_within_NML}), and that 
$\eta \sim 10$ is a reasonable assumption for the radio knots
(Section \ref{NML_Xrayknots}).
\\
\\[10pt]
\end{table*}

\begin{figure}[htb]
{\center
\includegraphics[width=.9\columnwidth]{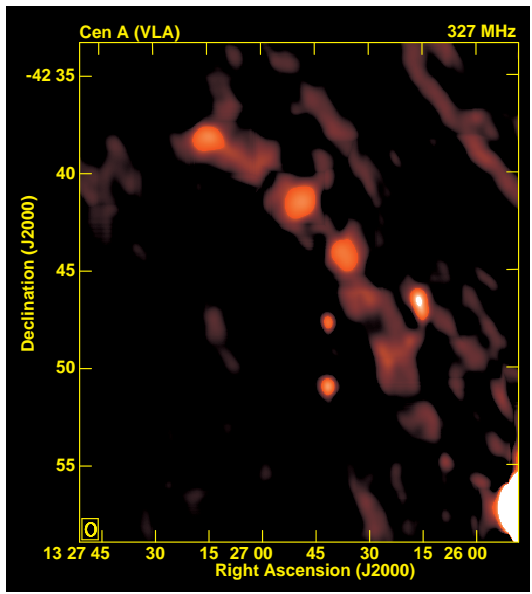}
\caption{A spatially filtered version of our 327-MHz image, which highlights
the compact features;  3 bright radio knots and the emission ridge which 
encloses them and extends $\sim 9\arcmin$ ($\sim 10$ kpc) to the SSE.  
(The three very compact
point sources S and W of the radio knots are background sources). Filtering
emphasizes emission on $\ltw 2.7\arcmin$ scales (Section \ref{Imaging}  and
Table \ref{ImageTable}). The restoring beam for this image is 
$40\arcsec \times 30\arcsec \times 0\arcdeg$.  The ridge of emission 
connecting the knots also connects to the linear M99 feature;  we believe
it is physically real.    We suspect other
lines to the NW of the knotty ridge are  imaging artifacts
(Section \ref{Obs_limitations} and Figure \ref{Fig:schematic_overall}),
also highlighted by the algorithm used to create this figure.
}
\label{Fig:radio_knots}
}
\end{figure}

Measurement of the compact knots and the ridge is complicated by the
smooth background emission on which they sit.  We therefore measured
the properties of the knots in our high-pass filtered image, which
removed most of the diffuse background emission, reducing the local
background to $\ltw$ 10 percent of the remaining peak 
knot brightness. Taking the remaining very local background and 
the ridge emission into account,  we measured the knot fluxes and 
sizes by examining slices across the knots (perpendicular to the
ridge; AIPS tasks SLICE and SLICEPL), and deconvolved the restoring 
beam of the image from the measured knot sizes.  The knots are slightly
resolved in our observations, with radii (defined as the half-power
half-width of the slice) $\sim 0.7$ kpc.  We used transverse slices 
across the ridge, in the high-pass-filtered image (Figure
\ref{Fig:radio_knots}), to estimate the full width 
of the ridge as $\sim3.4$ kpc.  Because we were not able to
differentiate clearly between the ridge and similarly scaled
filamentary structure that remained after high-pass filtering, we were
not able to obtain a robust measurement of the ridge's flux.

Our measured and derived results for the knots  
are presented in Table \ref{Table:NML_knot_numbers}.
Assuming the knots are synchrotron sources, we can derive minimum
pressures within the knots, as in Section \ref{Min_press_text}.
Our measured and derived quantities are 
given in Table \ref{Table:NML_knot_numbers};  our $p_{min}$
values  are summarized in
Table \ref{Table:pressures} and discussed in the next subsection.

We also checked whether the radio knots could be thermal bremsstrahlung sources.
If they have the characteristic $\nu^{-0.1}$ thermal spectrum, 
extrapolating from our 90-cm image predicts the knots should have
fluxes $\gtw 150-200$ mJy in the 20-cm image from M99. Because this
 is a factor of 2-3 larger than what is observed, 
it seems unlikely that the knots are thermal bremsstrahlung sources.

\subsection{Comparison: Radio and X-ray knots}  
\label{NML_Xrayknots}

\begin{figure}[htb]
{\center
\includegraphics[width=.85\columnwidth]{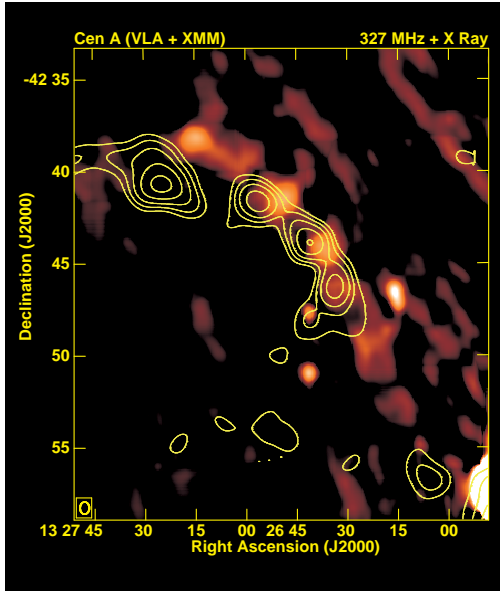}
\caption{Relationship between the X-ray knots (yellow contours;  from
K09) and the radio knots (background image, the high-pass-filtered image
from figure \ref{Fig:radio_knots}, 
beam $40\arcsec  \times  30\arcsec  \times 0\arcdeg$ ).  
X-ray data were taken by XMM-Newton
and  kindly provided by R. Kraft.  The inner radio and X-ray knots do not
coincide, but rather appear to alternate in a continuous, ridge-like
structure. }
\label{Fig:Radio_Xray}
}
\end{figure}

The knotty, radio-loud ridge shown in
 Figures \ref{Fig:best_overall} and \ref{Fig:radio_knots}
appears to be closely related to the similar strcture seen in soft X-rays
by K09. 
In Figure \ref{Fig:Radio_Xray} we show the relation between the two
structures; the X-ray and radio ridges sit almost on top of each other.
In the inner part of the ridge, the
X-ray and radio knots alternate, with the radio and X-ray knots offset
from each other by no more than the diameter of an individual knot 
($\ltw 1$ kpc).
Towards the furthest NE end of the structure,
the X-rays bend away from the radio ridge to the east, the separation between
the radio and X-ray knots increases, to as much as a few kpc.
As noted in Section \ref{TransRegion_overview}, this radio/X-ray 
ridge is an approximate demarcation between thermal, apparently quiescent
galactic ISM  east of the North Middle Lobe, and the nonthermal, 
radio-loud plasma which constitutes the North Middle Lobe and 
extends to the west of the ridge.

The radio-loud knots are comparable to the X-ray-loud knots in size and
pressure.  K09 cite radii $\sim 0.9\arcmin-1.5\arcmin$  for the X-ray knots;
we measure somewhat smaller radii, $\sim 0.6\arcmin - 0.7\arcmin$
 for the radio knots. K09  derive pressures $\sim 1.6-2.4 
\times 10^{-11}$\dyncm2 for the four knots within the X-ray ridge, under the
assumption that the X-ray knots are diffuse thermal emission.  
These
pressures are $\sim 20-30$ times higher than that of the galactic
ISM ($\sim 9 \times 10^{-13}$\dyncm2;  K09, also
 Section \ref{Diffuse_within_NML}).
We derive minimum pressures $\sim (1.1-1.2) \eta \times 10^{-12}$\dyncm2 for the
radio knots, under the assumption they are diffuse synchrotron emission.
If the pressure scale factor $\eta \sim O(10)$ -- as we argue holds for 
the Inner Lobe and North Middle Lobe region (Section
\ref{Min_press_text} and \ref{results_inner_lobes}) -- the pressure within the 
radio-loud knots is significantly higher than that in the nearby ISM and
the diffuse North Middle Lobe plasma, and comparable to or larger than 
that in the X-ray knots.

\subsection{Nature of the knots} 
\label{knot_nature}

We do not know the nature of the radio/X-ray knots, but can envision
three possibilites. 

\paragraph{Extant gas clouds}
It may be that both 
the radio and X-ray knots are simply pre-existing gas clouds 

in the region which -- for some reason -- 
are radio-loud, X-ray-loud, or both.  Although the two sets of knots
are not quite spatially coincident, the fact that they have similar
sizes and are similarly overpressure with respect to local ISM and to
the diffuse North Middle Lobe suggests that the radio 
knots and X-ray knots may be
related phenomena.  K09 proposed a pre-existing gas cloud model for
the X-ray knots and suggested that a fast plasma flow (such as a jet) from
the AGN has heated the clouds to X-ray temperatures. To extend this model 
to include radio emission one must add
additional physics, for instance shocks or turbulence to accelerate
locally  the relativistic  electrons which emit synchrotron radiation. 
Higher resolution radio and X-ray images could show bow shocks
from such a flow.

\paragraph{Multiphase emission-line clouds}
In a variant of the previous model, X-ray knots could be hot outer 
surfaces of multiphase clouds which also produce the ribbon 
of \hal and Far-Ultraviolet emission
in this region (Section \ref{TransRegion_overview}, also Paper 2). 
Similar clouds have been found elsewhere (e.g.,
the emission-line filaments in M87;  Sparks \etal 2010, 
 Werner \etal 2013), and are thought to have a cold interior (\hal emission),
a warm mid-region (Far-UV emission from CIV lines), and a hot outer layer 
(X-ray emission from Fe lines).  In Cen A, this idea is supported by the 
X-ray emission which Evans \& Koratkar (2004)
detected from the inner emission-line filament close to the galaxy.  
In addition, the X-ray clouds in Cen A
have clear internal structure, as might be expected for multiphase 
ISM clouds (Figures 3 and 5 of K09).  Once again,
radio emission is not an intrinsic part of this picture, but
would have to come from additional physics associated with the cloud's
existence and energization. High-resolution spectral imaging, at
X-ray and Ultraviolet wavelengths, could identify and characterize 
such clouds. 

\paragraph{Star formation sites}
An alternative possibility is that the radio and X-ray knots are sites of
ongoing star formation.  This would be consistent with the apparent
coincidence of the knots with the  extended string of emission-line
gas and young stars  in the same region (discussed in Paper 2).
Although the radio and X-ray knots are larger than 
``super star clusters'' studied elsewhere (typically $\sim 10$ pc across;
\eg, Melo \etal 2005 for M82), at least one very young cluster in the 
Antennae has a radius $\gtw $ 450pc (Whitmore \etal 1999). 
Interestingly, the luminosities of the 
knots\footnote{The radio knots have 
typical power $P_{1.4} \sim 1 \times 10^{27}$\ergps-Hz (using the fluxes
from Table \ref{Table:NML_knot_numbers} and assuming a spectral index
$\alpha = 0.7$ to convert to 1.4 GHz).  The X-ray knots have typical 
luminosity $L_x \sim 3 \times 10^{38}$\ergps (in the 0.5-2.0 keV band; 
K09).}  
agree quite well with the lower end of the radio-X-ray relation known to
hold for nearby star-forming systems (\eg, Ranalli \etal 2003, Mineo \etal
2014 and references therein). 
The proxy relations from Ranalli \etal and from Mineo et al. suggest that 
each knot is currently making stars at $\sim 0.1 M_{\sun}$/yr -- {\it if}
the knots are indeed extended star formation sites.  

In this picture, the X-rays are not simply from a diffuse gas but
rather  come from  a mix of X-ray binaries and hot plasma (heated by supernovae)
in the star-forming region.  The nature of the radio emission in this
relationship is again unclear, but it is likely to come from 
relativistic particles accelerated by supernova remnants, possibly supplemented
by radio jets driven out from  X-ray binaries.  If this picture holds,
and much of the X-rays or radio emission is not from diffuse plasma,
then our minimum pressure estimates of Section \ref{measurements_radio_knots}
would not apply.  High-resolution space observations, at Ultraviolet and
X-ray wavelengths, could identify star-formation sites.

\bigskip

\section{Energy Flow in the Transition Regions}
\label{Energy_Flow}

Although we have not found clear indications of collimated jets 
 within the North or  South Transition Regions of
Cen A, power from the inner galaxy must have moved through the
Transition Regions in order to create and maintain the Outer Lobes.
What can we say about the situation in those regions at present?

\subsection{Energy transport on large and small scales}
\label{Energy_throughput}

We can probe the situation in the Transition Regions by studying
energy flow into and through those regions.  
Because the Outer Lobes are on the order of a Gyr old, and the current
activity cycle of the AGN has lasted only  a few Myr, one might expect
the time-averaged power needed to create the OLs to be unrelated to
 the power currently being supplied by the AGN.  It turns out, however,
that these two powers are quite similar -- as we demonstrate here
and  summarize in  Table \ref{Power}.

\paragraph{AGN power based on  radio observations}
The energy flowing through the north jet in Cen A
 can be estimated from
high-resolution radio data.  Following Owen \etal (2000), the power
carried in a jet with radius $r_j$, pressure $p_j$, density $\rho_j$
speed $v_j $ and bulk Lorentz factor $\gamma_j$
is at least
as large as the power being advected out as internal energy:
$P_{j,int} \gtw 4 \pi r_j^2 \gamma_j v_j p_j$.
This quantity can itself be limited below by noting that the 
pressure within the jet is at least that given by the synchrotron
mininum pressure, $p_j \gtw p_{min,j}$.
To estimate the lower limit on $P_{j,int}$ we take 
take $r_j \sim 22$ pc,  $v_j \sim 0.5 c$,
from Goodger \etal (2010), and 
 use fluxes those authors measured for  the well-resolved knots.
 Assuming a spectral index $\alpha = 0.7$,
and guessing that $p_{min}$ across the jet $\sim 1/2$ that in 
the bright knots,
we get a lower limit on the power currently carried by the north
jet: $P_j \gtw  \eta \times 10^{43}$ \ergps.
 In Table \ref{Power}
we double that to estimate the total power currently being supplied,
to north and south, by the AGN. 

This estimate of the minimum jet power, today, depends on the unknown
pressure scale factor, $\eta$.  Although we have argued 
$\eta$ is of order 10   
 within the Inner Lobes and the diffuse North Middle Lobe -- based on
the likely cosmic ray composition of those plasmas -- we note that $\eta \gtw 1$
could describe an inner jet dominated by relativistic leptons.  If this
is the case, we have our most conservative lower limit for the jet power
currently being created by the AGN, $\gtw 2 \times 10^{43}$\ergps.

 \begin{table}[htb]
 \caption{Power estimates for Cen A}
 \label{Power}
 \begin{center}
 \begin{tabular}{c c c  }
 \hline
 \rule{0pt}{16pt} \hspace{-8pt }
 Epoch/region & Power$^a$ & Timescale \\
 &   (\ergps) & (Myr)
 \\[4pt]
 \hline
 \rule{0pt}{16pt} \hspace{-8pt} 
Present-day jets & 
 $\gtw 2  \times 10^{43}$ & today
 \\[2pt] 
Recent,  over IL life & 
 $\sim (2-6) \times 10^{43}$ & $\sim (1-2)$ 
 \\[2pt] 
Historical, over OL life$^b$
& $ \gtw (2 \!-\!5)  \times 10^{43}$ & 
  $\sim (400-600)/ \mu$ 
  \\[2pt]
 \hline
 \end{tabular}
 \end{center}
 $^a$ The total power to both sides of AGN,
assuming symmetric north and south outflows.\\
 $^b$ The cosine of projection angle between
 radio lobes and sky plane is $\mu$;  $\mu \sim 1/2$ for the inner
lobes, but is unknown for the outer lobes.  \\
\end{table}

\paragraph{AGN power based on X-ray observations}
The power needed to  create the inner radio lobes can  be estimated from
their impact on the galactic ISM.  
Croston \etal (2009) used X-ray data to characterize the shock which
the South Inner Lobe is driving into the local ISM.  That shock 
is moving in the sky 
plane at $2600$ \kmps relative to the ambient ISM.  
If the ISM is static, the age of
the South Inner Lobe is directly found from the shock speed and 5.5-kpc 
length of the South Inner Lobe (both quantities projected on the sky)
as  $\sim 2$ Myr.  If the
ISM is itself in motion (for instance in a fast wind, as we suggest
in Paper 2), the South Inner Lobe is younger, perhaps only $\sim 1$ Myr.  

Croston \etal (2009) used the shock measurements to estimate the power
put out by the AGN over the few-Myr lifetime of the Inner Lobes.  They used
the shock speed to find the work being done  by the inner
lobes expanding into the ISM, $p dV/dt \sim 7 \times 10^{42}$\ergps.  This
calculation is independent of the age of the South Inner Lobe,  
but is sensitive to the (unknown) projection angle of the South Inner 
Lobe.\footnote{Both VLBI and VLA data
  (\eg, Tingay \etal 2001, Clarke \etal 1992b) suggest the north jet
  and North Inner Lobe are pointed towards us, and the South Inner
  Lobe is pointed
  away from us, both at some intermediate angle (say $\sim
  45^{\circ}$) to the sky plane.}  If the South Inner Lobe 
does not lie in the sky plane, the true shock speed will 
exceed the measured value, and the
rate of work done will be correspondingly larger.  Croston \etal (2009) 
also used the same data to find the enthalpy content of the South Inner 
Lobe, and from that estimate the power needed to create the 
South Inner Lobe over an age $\tau_{IL}$ of Myr as $4 p V / \tau_{IL}
\sim 3 \times 10^{43}$ \ergps.  This estimate is insensitive to the
projection angle of the South Inner Lobe, but depends directly 
on the age;  if the South Inner Lobe is younger than 
2 Myr, more power is needed.  Taking
the uncertainties of both estimates into account, we collect the
South Inner Lobe-based power estimates as $\sim 1-3 \times 10^{43}$\ergps, and
double that (in Table \ref{Power}) to estimate the average power
supplied by the AGN on both sides over the past few Myr.

\paragraph{Power going  into the Outer Lobes}
The time-averaged power needed to create the outer lobes can be found
from their age and the energy required to expand them to their full size.
Following E14, the energy needed to build one of the 
Outer Lobes  is at least $E_{OL} \sim (2-3)\times 10^{59}/\mu$ erg.  
Unlike the Inner Lobes, we have no information on the projection angle
of the outer lobes, so we keep $\mu$, the cosine of that angle, as a variable.
The numerical range in $E_{OL}$ 
allows for uncertainty in the Outer Lobe plasma, which can be 
internally relativistic or can be dominated by thermal material.  If 
significant internal flows exist today, $E_{OL}$ is greater than this value.
The dynamic models presented by E14 show that the
age of the Outer Lobes is  $\tau_{OL} \sim (400\!-\! 600)/\mu$ Gyr;  
the range of numbers here reflects different possible  models.

Putting these results together, we find the time-averaged power $\langle
P_{OL} \rangle = E_{OL} / \tau_{OL}$ needed to grow each Outer Lobe, if
they contain no significant internal flows, 
is  $\sim (0.95 - 2.3) \times 10^{43}$ \ergps.
If significant flows exist within the lobes the necessary power is
larger.  As with the similar calculation for the Inner Lobes, this estimate
is nearly independent of $\mu$.  Multiplying this by a factor of two for the
two sides of the radio source, we find the average power put out by the AGN
over the $\sim$ Gyr lifetime of the source must have been
$ \gtw (1.9 \!-\! 4.6)  \times 10^{43}$  \ergps.

\subsection{Short-lived phenomena in and around the North 
Transition Region}
\label{Short_lived_NTR}

Although we do not know the detailed history of energy transport from the
inner galaxy -- steady or episodic -- we can
identify several phenomena 
which will soon decay or dissipate if they are not  maintained by some 
energy source active today in the North and South Transition Regions.

\paragraph{Diffuse plasma in the North Middle Lobe} 
\label{Overpressure_Diffuse}

We argued in Section \ref{Diffuse_within_NML} that pressure of
the diffuse radio-loud plasma in the North Middle Lobe is likely to be
several times larger 
than that of the quiescent ISM at a comparable distance from the galaxy.
Unless the North Middle Lobe is confined by some unseen mechanism 
(which we judge unlikely),
it cannot be a static structure, but should be expanding at approximately
its internal sound speed, $c_s$.  Because we don't know the thermal plasma
content of the North Middle Lobe, we cannot know $c_s$, but we can 
estimate its lower bound.  We take the  North Middle Lobe pressure as
$\gtw p_{min} \sim 4 \times 10^{-12}$\dyncm2 (from Table
\ref{Table:pressures}, taking $\eta \sim 10$ which we argued in 
Section \ref{Diffuse_within_NML} is probably the case for the North 
Middle Lobe as it
is for the Inner Lobes).  Recalling the hints of an X-ray
cavity in the region (Section \ref{TransRegion_overview}), we suggest the
density witin the North Middle Lobe
is no larger than the ambient ISM density, $\sim 10^{-3}$\cm3.
It follows that the local sound speed $c_s = (4 p / 3 \rho) \gtw 720$ \kmps
(using mean particle mass $0.6 m_p$).  Taking the characteristic radius
of the NML as $\sim 12$ kpc, we expect the North Middle Lobe to 
expand away in
in no more than $\sim 16$ Myr.

\paragraph{Radiative lifetimes in the NML}
\label{synch_lifetime}

The North Middle Lobe is detected by the Wilkinson Microwave Anisotropy
Probe (WMAP) at frequencies as high as 90 GHz
(Hardcastle \etal 2009).  Comparison of its flux there to the 5-GHz
flux reported by Junkes \etal (1993) shows a relatively flat spectrum
between 5 and 90 GHz.  Thus the high-frequency radiation cannot simply
be the ``tail'' of a fading electron distribution; we must ask if in
situ energization is needed.

Determining the synchrotron age of these electrons is
 difficult because we do not know the magnetic field
in the North Middle Lobe (recall that we do not assume 
that the minimum-pressure field,
from Section \ref{Min_press_text} and Appendix \ref{Min_pressure_App}, 
measures the true field value).  We can, however, 
find an upper limit to that age. Following E14, let the electrons lose
energy only by inverse compton scattering  (Stawarz \etal 2006 show
that the CMBR dominates galactic starlight at the distance of the North 
Middle Lobe), until
they find themselves in a strong field within the North Middle Lobe. 
 This gives us
the largest possible synchrotron life. If we guess
that strong field $\sim 10 \mu$G (about twice the field at which magnetic
pressure balances local ISM pressure -- 
as might be appropriate for shock-enhanced fields within the NML) the
maximum lifetime for electrons radiating at $\nu$ 
is $\sim 180 \nu_{GHz}^{-1/2}$ Myr (from E14).   This shows that
electrons in the North Middle Lobe radiating at $\sim 1-10$ GHz  do not require
{\it in situ} energization, but electrons radiating at 90 GHz will fade in 
no more than $\sim 20$ Myr without re-acceleration.


 \begin{table}[htb]
 \caption{Timescales in the Transition Regions}
 \label{Short_Times}
 \begin{center}
 \begin{tabular}{c c c  }
 \hline
 \rule{0pt}{16pt} \hspace{-8pt }
 Object &  Timescale & Notes\\[-2pt]
 \\[4pt]
 \hline
 \rule{0pt}{16pt} \hspace{-8pt} 
Diffuse NML$^a$ & $\ltw 16$ Myr & overpressure expansion
 \\[2pt] 
Synchrotron aging & $\ltw 20 $ Myr & at/above $\sim 90$ GHz
 \\[2pt] 
OL re-energization & $\ltw 30$ Myr & turbulent decay  in OLs
 \\[2pt] 
Radio/X-ray knots$^b$ & $\sim 0.5\!-\!3$ Myr &  if diffuse gas clouds
 \\[2pt] 
Radio/X-ray knots$^c$ & $\sim 10 $ Myr & if star-forming regions
 \\[2pt] 
 \hline
 \end{tabular}
 \end{center}
$^a$ Assumes the diffuse North Middle Lobe is at higher pressure
than its surroundings (Section  \ref{Diffuse_within_NML}).
\\
$^b$ Applies if the radio and X-ray knots are diffuse,
unconfined plasma (Section \ref{knot_nature}).
\\
$^c$  Applies if the X-ray knot emission comes from massive
young stars (Section \ref{knot_nature}).
\end{table}

\paragraph{Re-energization of the Outer Lobes}
The Outer Lobes are much older than the radiative lifetime of
their relativistic electrons, especially those detected in $\g$-rays
(E14, also Section \ref{Intro:large_scales}).  They must, therefore,
be frequently re-energized.  E14 argued this must happen
at least every few tens of Myr to maintain the internal turbulence which 
can re-accelerate the electrons. Because the Transition Regions
are the only conduit to the Outer Lobes, it follows that some form
of energy transport must have permeated {\em both} the North and South
Transition Regions  at least as recently as $\sim 30$ Myr ago.

\paragraph{Radio and X-ray knots: diffuse gas clouds}
In Section \ref{knot_nature} we noted that the radio and X-ray knots
could be either diffuse gas clouds or regions of star formation.
If the knots are diffuse gas clouds, 
they are strongly overpressured, $\gtw 20 p_{ISM}$.  Unless they are
somehow confined, they will dissipate quickly. 
As K09 discuss, if the
 X-ray knots are unconfined structures with size $\sim 1$ kpc
and  temperature $T \sim 0.5$ keV, they 
will expand at their adiabatic sound speed in only $\sim 3$ Myr. 
For the radio knots, we know the radius ($\sim 0.7$ kpc,
Section \ref{results_radio_knots}), but -- as with the 
diffuse North Middle Lobe -- we do not know their internal sound speed.
We therefore find its lower bound by taking the smallest
likely pressure and the largest likely density.
Let a radio-loud cloud 
contain plasma  at the local ISM density, $\sim 10^{-3}$\cm3, with 
pressure $\gtw 10^{-11}$\dyncm2 (Table \ref{Table:pressures}, taking
$\eta \sim 10$).  Its sound speed is thus $\gtw 1300$ \kmps.
Such a cloud, if unconfined, will last no more $\sim 0.5$ Myr.
Thus, if the radio and X-ray knots are separate clouds within
the weather system, both will dissipate in no more than few Myr unless
they are somehow confined.

\paragraph{Radio and X-ray knots: clumps of young stars}
We also noted in Section \ref{knot_nature} that 
the X-ray and radio knots may be  star-forming regions. If this is the
case, the region cannot be age-dated by the 
the high knot pressures.  If the radio and X-ray emission 
comes from diffuse
ISM witin the region, self-gravity of the star cluster and/or wind outflow
supplied by the star formation can maintain the high pressures over a longer
timescale.  However, X-ray emission from star-forming regions 
can also come from X-ray binaries.  Hard X-ray emission comes from
high-mass binaries, which only last  $\sim 10$  Myr.  
Because the knots have only been observed 
at soft X-ray energies (K09), and because soft X-rays can have origins other 
than high-mass stars, we cannot with certainty state that the radio/X-ray
knots contain young stars. However,
observational evidence for such young stars elsewhere within
the weather ribbon (\eg, Graham 1998, Mould \etal 2000) suggests that
similarly young stars might exist within the X-ray/radio knots {\it if} 
they are star-forming regions.

\paragraph{Other short-lived weather in the North Transition Region}
The above discussion shows that several features related to the radio
and X-ray emission in the Transition Regions probably
require re-energization
on timescales of a few to a few tens of Myr.  We summarize these features
in Table \ref{Short_Times}.  In addition, we show in  
paper 2 that  a trail of emission line filaments and young stars
stretches $\sim$ 35 kpc outward from the galaxy, along the southeast
edge of the North Transition Region.  
Gas dynamics, cooling times, and stellar ages in this system
all indicate timescales no longer than $\sim 10$ Myr.

\subsection{Energy flow through the Transition Regions}
\label{Energy_flow_in_TRs}

The power estimates in Table \ref{Power} show that the 
current output of the AGN is sufficient to create and maintain the
large-scale radio source, if provided consistently over the $\sim$ Gyr
lifetime of the  source.  This power must have moved outwards,
through the North and South Transition Regions, as the Outer Lobes
grew to their present size.  If the North Transition Region
 contains such a flow at present,
that flow can potentially provide the driver needed for the short-lived
phenomena in the region.  Although we do not know the 
history of the AGN, we can consider different
possibilities.

\paragraph{An episodic AGN with a low duty cycle?}

The youth of the Inner Lobes (only $\sim 1\!-\!2$ Myr old) 
suggests that  power from the AGN has
not been steady, but rather has fluctuated with time.  How long can
the AGN have been ``off''?
As an example, say the AGN has been quiet for $\gtw 30$ Myr (the longest
lifetime in Table \ref{Short_Times}). If remnants of plasma
 from the previous active cycle of the AGN
have been coasting  at 3000 \kmps (the likely deprojected
advance speed of the end of the South Inner Lobe; 
Section  \ref{Energy_throughput}), they
would have travelled $\gtw 100$ kpc, well past
the Transition Regions.  Therefore, any
material left in the Transition Regions 
today would be nearly static, and unable to maintain the
short-lived phenomena we see in the North Transition Region.
We conclude that such a long ``down time'' for the AGN is unlikely.

\paragraph{An AGN with a high duty cycle?}

It may also be that the AGN is  ``on'' for a large fraction of the time,
but with a time-variable outflow. It may  drive out collimated jets, or 
a broader diffuse flow, or it may alternate between these two states. 
 If this is the case, plasma in the Transition
Regions is probably still in motion,  because the flows 
 from the previous active cycle (before the current one that has 
created the Inner Lobes)  have not yet moved through the regions.  
We find this  model attractive -- because it does not
require us to be catching Cen A at a special time -- but we cannot prove
it is the case. 

We note that the starting plumes which we see as today's Inner Lobes
might evolve in two different ways as they move through the Transition
Regions. They may continue to be identifiable as jets, similarly to
the inner jets in 3C219 (Clarke \etal, 1992a).  If this is the case,
Transition Regions will look quite different after $\sim 15$ Myr when
the new jets have reached them.  Alternately, the Inner Lobes might
evolve, via instabilities, into a broader, diffuse flow or turbulent
plume, perhaps similar to the rapidly broadening plumes in M84 (\eg,
Laing \etal 2011). When these diffuse plumes reach the Transition
Regions, they would not necessarily be identifiable as a confined jet,
and  the Transition Regions would look about the same in $\sim 15$
Myr as they do now.

\paragraph{No jets in the Transition Regions?}

Although both our observations and those of
M99 detect linear structures near the SE edge  
of the North Transition Region, which may connect to each other,
we do not believe these 
can be the main source of power for the North Transtion Region
or the North Outer Lobe.  Our reasoning is as follows.

$\bullet$
Classic jets in well-studied radio galaxies are narrow, well-collimated,
continuous 
outflows which begin at the AGN and carry mass and energy to the large-scale
radio lobes.  Such a jet clearly exists in the North Inner Lobe of Cen A.
 It can be traced from the core to $\sim 5$ kpc north,  where 
it bends toward the  west and appears to disrupt (Figure \ref{Fig:InnerLobes},
also Clarke \etal 1992b). The North Inner Lobe has a sharp northern edge, with
 no sign of any ``jet breakout'', 
 which  would be expected if the linear M99 feature were a continuation
of the north jet flow currently powering the North Inner Lobe.

$\bullet$ There is no sign of a jet in the South Inner Lobe, although power
is very likely being supplied to it at present (evinced by shock expansion,
Section \ref{Intro:small_scales}). 
Furthermore, there is no sign of any collimated flow past the South Inner
Lobe, continuing into and through the South Transition Region, 
although ongoing power is needed to keep the both the South and North
Outer Lobes  shining in $\gamma$-rays (Section  \ref{Intro:large_scales}). 

$\bullet$
Although the diverse short-lived phenomena in the 
North Transition Region  need ongoing energization, there is 
no clear connection between the
narrow linear M99 feature, or the knotty ridge we detect at 90 cm, 
  and the broader diffuse North Middle Lobe. As we
show in paper 2, the linear M99 feature is also offset from many of
the emission-line clouds and young stars in the region,
 which also need ongoing drivers. It therefore seems unlikely that this
faint, narrow feature can drive the entire system. 

$\bullet$
By contrast with a jet -- a narrow, collimated
outflow emanating from the AGN -- we envision a ``wind'' as a more diffuse 
structure.  It can be broader, less well collimated, and
may come from a larger 
region of the galactic core.  Well-studied 
examples (the solar wind, or winds from starburst galaxies)
 tell us that  wind flows can be inhomogeneous, 
 with internal shear surfaces, shocks, and localized high-speed streams.  
We suspect that the knotty ridge we observe at 90 cm, as well as the 
 linear M99 feature,
 indicate similar inhomogeneities in a diffuse wind moving through
the North Transition Region of Cen A.

\section{Final Words}
\label{The_Last_Section}

Our new findings, reported in this paper, change and enhance 
our picture of the North and South Transition Regions in 
Cen A.  

$\bullet$
{\em We see no indication of a large-scale jet in either Transition Region.}
 We do not find any evidence of a radio jet (which we define as a
continuous, highly collimated, plasma flow from the AGN) in the middle
regions of Cen A.  Our observations show the diffuse radio emission known
as the  North Middle Lobe contains a bright, knotty, 
emission ridge close to the southeast edge of the region.
We suggest that the linear feature reported by M99, and 
interpreted by many authors as a  large-scale jet, is 
 part of this radio ridge.  However, the true nature of this region
remains elusive, due to technical limitations of existing images.
We look forward to future observations with next-generation telescopes which
will clarify the structure of the regions.

$\bullet$
{\em Energy must have been moving through both Transition Regions very
recently.}  
The  energy provided by
the current level of nuclear activity, if continually supplied,
is sufficient to create and maintain the Outer Lobes over the $\sim$ Gyr
lifetime of Cen A.  
Although we have no direct evidence of plasma flow within
the Transition Regions right now, several short-lived structures
in the NTR indicate that something must have energized those 
phenomena on the order of $\sim 10$ Myr ago, if not more recently.
The need for ongoing re-energization of 
the Outer Lobes also shows that energy must have moved through both
Transition Regions no more than $\sim 30$ Myr ago.

$\bullet$
{\em Previous models of the North Middle Lobe are inadequate.} 
Many of the models suggested to explain the North Middle Lobe  
cannot work,
because they depend on the existence of a collimated jet 
passing through the region now.  Models which
depend on the North Middle Lobe being a slowly rising bouyant 
bubble have trouble explaining the  short-lived phenomena in 
the North Transition Region and the need for
rapid re-energization of both Outer Lobes.  

$\bullet$
In Paper 2 we return
to the question of energy flow and short-lived phenomena
 within the Transition Regions. We show that a wind is expected in
Cen A/NGC 5128, discuss observational signatures of such a wind,
 and argue that such a wind 
 naturally explains the enhanced radio emission and
other weather phenomena seen in the North Transition Region.

\bigskip

\begin{acknowledgements}

We are very grateful to R. Morganti, N. Junkes,  and  R. Kraft  
for sharing their data with us. 
We thank E. Greisen for consistently friendly software 
support throughout this project.  We appreciate thoughtful comments
from the referee  which have improved this paper. 
SGN thanks the NRAO in Socorro, NM, for hospitality during 
major parts of this work.

In this work we have made extensive use of both NASA's Astrophysics 
Data System (ADS; hosted by the High Energy Astrophysics Division 
at the Harvard Smithsonian Center for Astrophysics), and the 
NASA/IPAC Extragalactic Database (NED; operated by the Jet 
Propulsion Laboratory, California Institute of Technology, 
under contract with NASA)

\end{acknowledgements}

\bigskip

\begin{appendix}

\section{Minimum pressure analysis}
\label{Min_pressure_App}

The energy
content in relativistic particles and magnetic fields is a
 key factor in the physics of synchrotron sources. 
Because synchrotron power depends
on the product of the two, these cannot be
determined uniquely from the data, 
One way to separate particles from fields
is to minimize the total pressure, $p_{rel} + p_B$, subject to the constraint
that the volume emissivity is known (e.g. Burns, Owen \& Rudnick 1979).
Details matter here; 
variants of this method found in the literature can be confusing.
  In this appendix we describe the specific calculations we present in
Section \ref{Results_section}.

To implement this, we must connect the 
observed synchrotron power to the total pressure in 
relativistic electrons.  If the electron energy distribution $n(\gamma)$ is 
defined over some range $\gamma_1 < \gamma < \gamma_2$, the pressure is
\be
p_{rel, e} = { 1 \over 3} 
\int_{\gamma_1}^{\gamma_2} n(\gamma) \gamma m c^2 d \gamma
\label{pressure_int_1}
\ee
We assume the usual electron power law,  $n(\gamma) \propto \gamma^{-s}$,
which produces a power-law synchrotron spectrum, 
$\epsy(\nu) \propto \nu^{-\alpha}$,  where the spectral index
$\alpha = (s-1)/2$. 
The choice of limits  in equation (\ref{pressure_int_1}) 
matters because $p_e$ is dominated by electrons
close to the low-energy cutoff, but radio observations generally do not
extend to low enough frequencies to determine this cutoff. 

Many authors (e.g, Pacholczyk
1970, Burbidge 1956) choose the range of electron energies 
corresponding to a
specific range of observed radio frequencies, $\nu_1 < \nu < \nu_2$,
 and assume no electrons
radiate outside this range. In principle this
frequency range should be chosen based on the broad-band spectrum of a
given source, but in practice is typically parameterized as, say, 10 MHz to
10 GHz.  Because the electron energy is related to the 
emitted frequency by $\nu =
a \gamma^2 B$ (for  $a = 3 e / 4 \pi m c$), the integral in equation 
(\ref{pressure_int_1}) can be written
\be
p_{rel, e} = { 1 \over 3} \int_{(\nu_1 / a B)^{1/2}}^{(\nu_2/a B)^{1/2}}
 n(\gamma) \gamma m c^2 d \gamma 
\label{pressure_int_2}
\ee
With this, the electron pressure, $p_e$, 
is connected to the synchrotron emissivity,
$\epsy(\nu_o)$ at some observed frequency $\nu_o$, as
\be
p_{rel, e} = { 1 \over 3} { A^{BP} \over B^{3/2}} ~; \quad
A^{BP} =  {\epsy (\nu_o) \nu_o^{\alpha} \over
( 2 \alpha -1)} { 2 c_1^{1/2} \over c_2}  
{\left[ 1 - ( \nu_2 / \nu_1)^{-(2 \alpha -1)/2} \right] 
\over ( \sin \theta)^{3/2} \nu_1^{( 2 \alpha -1)/2}}
\label{elec_press_trad}
\ee
Here $\sin \theta$ is the pitch angle (between particle velocity and 
magnetic field);  $B \sin \theta$ 
can also be thought of as the projection of the B field on the sky plane.
The numerical factors  $c_1, c_2$ are 
 given in Pacholczyk (1970). In our calculations we take $\nu_1 = 10$ MHz,
and choose $\alpha = 0.7$.  For this spectral index, the derivations are only
weakly sensitive to $\nu_1$: $A^{BP} \propto \nu_1^{-0.2}$.  We also assume 
$\left[ 1 - ( \nu_2 / \nu_1)^{-(2 \alpha -1)/2} \right] 
/ ( \sin \theta)^{3/2}  \simeq O(1)$.

An alternative approach was proposed by Myers \& Spangler (1985).
 They argued that 
the low-energy electron cutoff is set by the physics of the source, not the
happenstance of our observations.  They treat the low-energy cutoff in the
electron energies, $\gamma_1$, as an input parameter, 
and retain the form (\ref{pressure_int_1}) in subsequent analysis.  
To implement this approach, we  connect $p_{rel,e}$ to the emissivity as
\be
p_{rel, e} = { 1 \over 3} { A^{MS} \over B^{\alpha + 1}} ~; \quad
A^{MS} =   { \epsy (\nu_o) \nu_o^{\alpha} 
\over ( 2 \alpha -1) 4 \pi c_5(\alpha)}
{ ( 2 c_1)^{- \alpha}
\over 
( m_e c^2)^{(2 \alpha -1)} } 
{ \left[ 1 - ( \gamma_2 / \gamma_1)^{-(2 \alpha -1)} \right] 
 \over \gamma_1^{( 2 \alpha -1)} ( \sin \theta )^{\alpha +1}}
\label{elec_press_MS}
\ee
where $c_5(\gamma)$ is another constant from Pacholczyk.
In our calculations we take $\g_1 = 10$, and $\alpha = 0.7$.  
With this spectral index, our calculations are again only weakly dependent on
$\gamma_1$:  $A^{MS} \propto \g_1^{-0.4}$.  We  also 
assume $\left[ 1 - ( \gamma_2 / \gamma_1)^{-(2 \alpha -1)} \right] 
/ ( \sin \theta )^{\alpha +1} \simeq O(1)$.

To evaluate $A^{BP}$ or $A^{MS}$, we derive the
 synchrotron volume emissivity, $\epsy (\nu_o)$ form 
the observed flux  at $\nu_o$ by $\epsy  (\nu_o) = 4 \pi D^2 F(\nu_o)/V_{emit}$, 
inferring the
emitting volume $V_{emit}$ from observations. One immediate uncertainty
appears here:  the volume inferred from an image, $V_{obs}$,
  may be incompletely filled, as $V_{emit} = \phi V_{obs}$, 
with filling factor $\phi < 1$.  A second uncertainty
is how to connect $p_{rel,e}$ to the total plasma pressure, $p_{plasma}$.  
This quantity can 
include relativistic baryons and, possibly, other contributions
such as thermal plasma.  We follow the usual approach by writing
$p_{plasma} = (1+k) p_{rel,e}$.  We retain both 
 $\phi$  and $k$ as unknown parameters; in our calculations (below)
they combine as $(1+k)/\phi$.

Either equation (\ref{elec_press_trad}) and (\ref{elec_press_MS}) can be
used to minimize the total pressure of the magnetized plasma,
\be
p_{rel} = p_{plasma} + p_B = ( 1 + k) p_{rel, e} + {B^2 \over 8 \pi}
\label{min_p_expression}
\ee
For the BP method,  
minimizing equation (\ref{min_p_expression}) with respect
to $B$ leads to
\be
p_{min}^{BP} = 
{ 7 \over 24} \left[ 2 \pi A^{BP} {( 1 + k) \over \phi} \right]^{4/7}
= { 7 \over 24} \left[ 2 \pi A^{BP}] \right)^{4/7} \eta^{BP}
\label{minP_BP}
\ee
where we have defined 
\be
\eta^{BP} = \left[ ( 1 + k ) \over \phi\right]^{4/7}
\label{etaBP}
\ee
Also useful relations for the BP case include
\be
 p_{plasma}  = { 4 \over 3}p_B ~; \quad 
B_{min~p}^{BP} = \left[ 2 \pi A^{BP} \right]^{2/7} 
\left[ \eta^{BP} \right]^{1/2}
\label{useful_relations_BP}
\ee

For the MS method,  minimizing equation (\ref{min_p_expression}) with respect
to $B$ leads to
\be
p_{min}^{MS} 
= { 3 + \alpha \over 8 \pi ( 1 + \alpha)} \left[ { 4 \pi \over 3} ( 1 + \alpha)
A^{MS} {(1 + k) \over \phi} \right]^{2 / ( 3 + \alpha)}
= { 3 + \alpha \over 8 \pi ( 1 + \alpha)} \left[ { 4 \pi \over 3} ( 1 + \alpha)
A^{MS}  \right]^{2 / ( 3 + \alpha)} \eta^{MS}
\label{minP_MS}
\ee
where 
\be
\eta^{MS}= \left[ {( 1 + k) \over \phi } \right]^{2 /(3 + \alpha)}
\label{etaMS}
\ee
We again have the useful relations for the MS case,
\be
 p_{plasma}  = { 2 \over 1 + \alpha} ~p_B ~; \quad
B_{min~p}^{MS} = \left[ { 4 \pi \over 3} ( 1 + \alpha) A^{MS} 
\right]^{1/(3+\alpha)}  \left[ \eta^{MS}\right]^{1/2}
\label{useful_relations_MS}
\ee

For spectral index $\alpha = 0.7$, our choice in this paper, 
the numerical values of $\eta^{BP} = \left[ ( 1 + k ) / \phi\right]^{0.57}$
 and $\eta^{MS} = \left[ {( 1 + k) / \phi } \right]^{0.54}$ are quite
similar.  In discussion in the paper we drop the ``MS'' and ``BP'' superscripts
and use the approximate $\eta \sim \left[ (1+k) /\phi\right]^{1/2}$ 
in  numerical  estimates.

\end{appendix}

\clearpage


\section{Regions where fluxes and areas were measured - OnLine-Only Figures}



\begin{figure}[htb]
{\center
\includegraphics[width=0.95\columnwidth, trim = 2cm 6cm 2cm 6cm, clip=True]{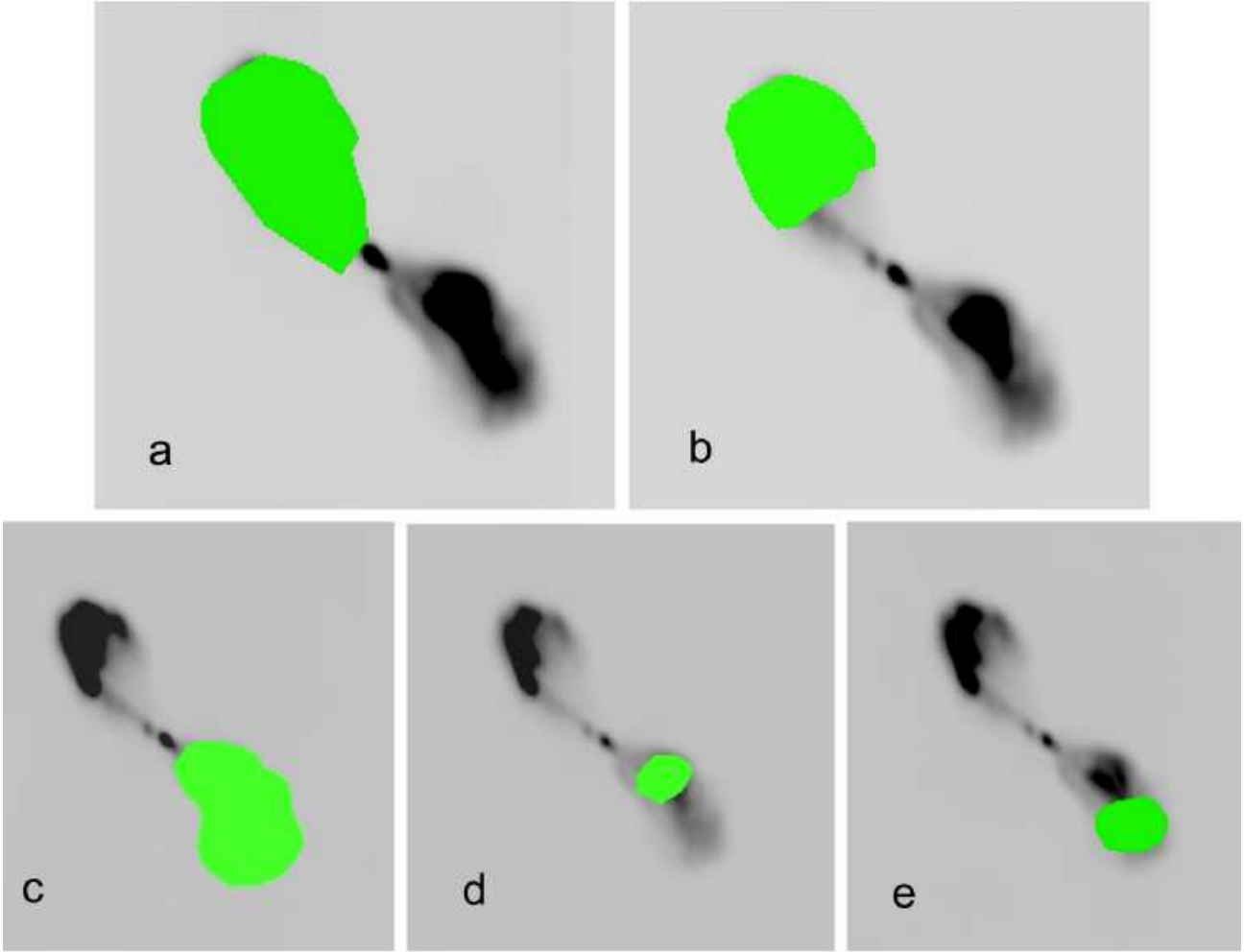}
\caption{Regions in the inner lobes where fluxes were measured, 
as given in Table \ref{Table:IL_NML_numbers}.  Measurements were done  in images made
by combining VTESS models with residual flux images, as discussed in Section \ref{Imaging}.
All images were convolved with an elliptical restoring beam 28\arcsec $\times$ 17\arcsec,
oriented at position angle 11$^{\circ}$.
~~a) The full northern inner lobe.  ~~b The outer part of the northern inner lobe
~~c The full southern inner lobe; 
~~d and ~~e The inner and outer parts of the southern inner lobe.
{\high (ONLINE FIGURE only)} 
}
\label{Fig:green_blotches_ILs}
}
\end{figure}

\begin{figure}[htb]
{\center
\includegraphics[width=0.95\columnwidth]{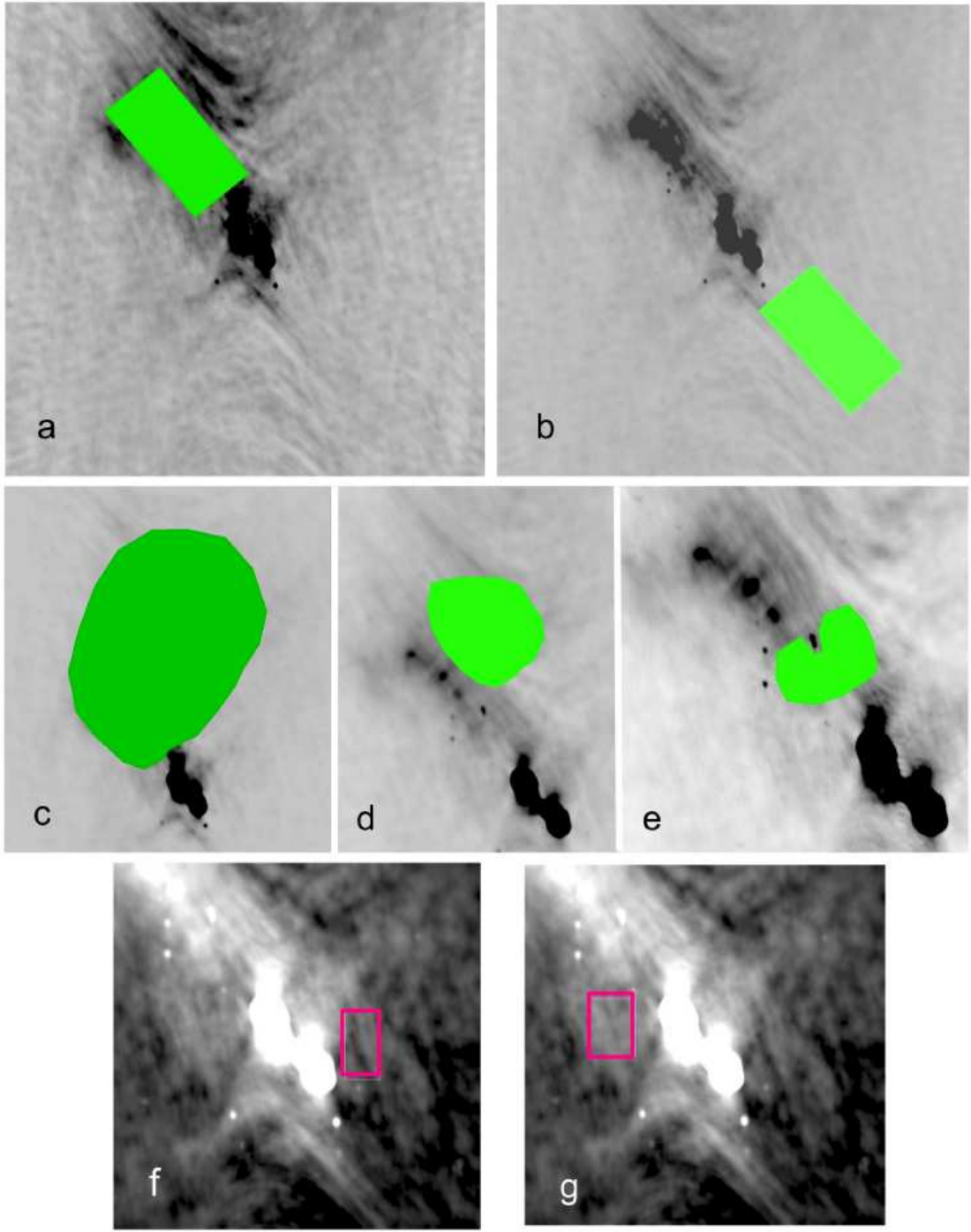}
\caption{ Regions outside the inner lobes where fluxes were measured, as in 
Table \ref{Table:IL_NML_numbers}.  Images were smoothed 
by different size restoring beams; all restoring beams were oriented N-S.   
~~a) The North Middle Lobe, region measured by Kraft (2009). 
~~b) A comparable area where a South Middle Lobe might be expected.  
The images in both  ~a) and ~b) have been convolved with
an elliptical restoring beam 35\arcsec $\times$ 25\arcsec.
~~c) Full NML, in image convolved by restoring beam  
70\arcsec $\times$ 60\arcsec.
~~d) Diffuse emission Region 3 in Table \ref{Table:IL_NML_numbers}, 
shown on image
with restoring beam  40\arcsec $\times$ 30\arcsec. 
~~e) Diffuse emission Region 1  in Table \ref{Table:IL_NML_numbers}, 
shown on image
with restoring beam  28\arcsec $\times$ 17\arcsec. 
~~f) and ~g) Local regions \#1 and \#2, near but outside Inner Lobes in Table 
\ref{Table:IL_NML_numbers}, both shown on images
with restoring beam  35\arcsec $\times$ 25\arcsec. 
{\high (ONLINE FIGURE ONLY)}
}
\label{Fig:green_blotches_MLs}
}
\end{figure} 

\end{document}